\documentclass[12pt]{article}
\usepackage{geometry}
\usepackage{amsthm,amsmath,latexsym,amssymb,amsfonts,amscd}
\usepackage{graphics,lscape,fancyhdr,array,stmaryrd,euscript,wrapfig}
\usepackage{empheq,wrapfig}
\usepackage{verbatim,slashed}
\numberwithin{equation}{section}
\usepackage{hyperref,setspace}
\usepackage{tikz-cd}
\usepackage{mathrsfs}
\usepackage[numbers,sort&compress]{natbib}
\usepackage[nottoc]{tocbibind}

\newcommand{\pl}{\partial}

\newcommand{\fud}[2]{{}^{#1}{}_{#2}\,}
\newcommand{\fdu}[2]{{}_{#1}{}^{#2}\,}

\newcommand{\besubeqs}{\begin{subequations}}
\newcommand{\esubeqs}{\end{subequations}}

\begin{document}
\pagenumbering{gobble}
\hfill
\vskip 0.01\textheight
\begin{center}
{\Large\bfseries 
Massive spin three-half field \\ [5pt] in a constant electromagnetic background }

\vspace{0.4cm}

\vskip 0.03\textheight
\renewcommand{\thefootnote}{\fnsymbol{footnote}}
William \textsc{Delplanque}, 
Evgeny \textsc{Skvortsov}\footnote{Research Associate of the Fund for Scientific Research -- FNRS, Belgium}\footnote{Also at Lebedev Institute of Physics.}
\renewcommand{\thefootnote}{\arabic{footnote}}
\setcounter{footnote}{0}
\vskip 0.03\textheight

{\em Service de Physique de l'Univers, Champs et Gravitation, \\ Universit\'e de Mons, 20 place du Parc, 7000 Mons, 
Belgium}

\end{center}

\vskip 0.02\textheight

\begin{abstract}
Massive higher-spin fields are difficult to introduce consistent interactions, including electromagnetic and gravitational ones which are clearly exhibited by (non-elementary) higher-spin particles in nature. We construct an action that describes consistent interactions of massive spin three-half field with a constant electromagnetic background. We also work out the relation to the chiral approach.
\end{abstract}

\newpage
\tableofcontents
\newpage
\section{Introduction}
\label{sec:}
\pagenumbering{arabic}
\setcounter{page}{1}

Since 1939 \cite{Wigner:1939cj} all particles, be elementary or not, must fall into Wigner's classification that, excluding some exotic cases, see e.g. \cite{Bekaert:2006py,Bekaert:2017khg}, assigns two parameters to every particle in $4d$ --- spin and mass. The two parameters are associated with unitary irreducible representations of the Poincar{\'e} algebra in $4d$ and provide us with a list of free elementary systems that are consistent with quantum mechanics and Poincar{\'e} symmetry. A fundamental question is which multiplets of particles admit consistent classical and, then, maybe quantum theories. Very few options are available at present with varying degrees of (in)consistency at the quantum level, e.g. gauge theories, (super)gravities, string theories, massive (bi)gravities \cite{Bergshoeff:2009hq,deRham:2010kj,Hassan:2011zd,deRham:2014zqa} and a handful of higher spin gravities \cite{Bekaert:2022poo}. There is also a great disparity between low spin and higher spins. 

Facts and reality tell us that there are plenty of massive higher-spin particles that are non-elementary, e.g. hadrons or nuclei, with many of the latter being stable. Whenever the gravitational and electromagnetic fields are small enough (which is not hard to arrange) the particles can effectively be treated as elementary and are known to exhibit electromagnetic and gravitational interactions. It then comes as a surprise that there does not exist a simple theoretical gadget to construct such (effective) interactions that maintain the correct number of degrees of freedom when interactions are introduced, which can be thought of as a generalized Boulware-Deser problem \cite{Boulware:1972yco} or not unrelated Velo-Zwanziger one \cite{Velo:1969bt}.

Theoretically, massive higher-spin fields can be described by symmetric (gamma)-traceless (spin)-tensors $\Phi_{\mu_1...\mu_s}(x)$, \cite{Fierz:1939ix,Singh:1974qz,Singh:1974rc}. The price for the manifest Lorenz invariance is that $\Phi_{\mu_1...\mu_s}$ contains more components than the number of physical degrees of freedom, $2s+1$. The redundant components are to be eliminated via the transversality constraint $\pl^\nu \Phi_{\nu \mu_2...\mu_s}=0$, whose Lagrangian implementation requires a host of auxiliary fields \cite{Fierz:1939ix,Singh:1974qz,Singh:1974rc}. It is a challenge to prevent these unphysical degrees of freedom from propagation when interactions are turned on, which requires a tedious analysis of Hamiltonian constraints. A more streamlined approach is to enlarge the field content even more, see e.g. \cite{Pashnev:1989gm,Zinoviev:2001dt,Zinoviev:2006im,Zinoviev:2008ck,Zinoviev:2009hu,Zinoviev:2010cr,Buchbinder:2012iz}, as to introduce the Stueckelberg-like gauge symmetries. Consistent interactions have to be gauge invariant, at the very least, but certain additional assumptions on the number of derivatives are needed. 

The chiral approach has been proposed recently in \cite{Ochirov:2022nqz}. The idea is to eliminate the need for the transversality constraint by introducing a chiral field $\Phi_{A_1...A_{2s}}$ in $(2s+1)$-dimensional representation $(2s,0)$ of the Lorentz algebra $sl(2,\mathbb{C})$. Since there are no redundant components, interactions are easy to introduce. However, the discrete symmetries, most importantly the parity, are difficult to implement. Nevertheless, it was demonstrated in \cite{Delplanque:2024enh} that chiral and the usual (worth calling it symmetric) approaches are equivalent up to spin-two. Other ideas to introduce interactions include covariant techniques \cite{Buchbinder:2005ua,Buchbinder:2007ix,Kaparulin:2012px,Kazinski:2005eb} and the light-cone gauge \cite{Metsaev:2005ar,Metsaev:2007rn,Metsaev:2022yvb}.

One easy-to-formulate open problem is how to make massive higher-spin fields propagate on electromagnetic and gravitational backgrounds. A subproblem, which we address in the present paper, is to restrict to the constant electromagnetic background. Massive spin-one fields interacting with an external electromagnetic field can be obtained via the Brout-Englert-Higgs mechanism. Therefore, the first nontrivial case is that of the massive spin three-half. The story of the spin three-half has been quite long and often negative, see e.g. \cite{Rarita:1941mf,Johnson:1960vt,Velo:1969bt,Zinoviev:2009hu,Deser:2000dz,Deser:2001dt,Buchbinder:2014hfa,Khabarov:2021djm,Benakli:2023aes,Benakli:2022edf,Benakli:2021jxs} and references therein/thereon. For the case of genuine higher-spin fields see e.g. \cite{Klishevich:1998sr,Klishevich:1998ng,Klishevich:1998yt,Zinoviev:2008ck,Porrati:2010hm,Buchbinder:2012iz,Buchbinder:2015uea}.

For constant electromagnetic backgrounds some results were obtained from string theory, see e.g. \cite{Argyres:1989qr,Argyres:1989cu,Klishevich:1998sr,Porrati:2010hm,Benakli:2022edf,Benakli:2023aes,Benakli:2021jxs}. However, the space-time dimension is fixed and cannot be dialed to $4$ easily \cite{Porrati:2010hm}, which is the main case of interest. Instead, one can compactify to $4d$ to find that different states mix when interactions are turned on \cite{Porrati:2010hm,Benakli:2022edf,Benakli:2023aes,Benakli:2021jxs}, which also occurs before compactification \cite{Klishevich:1998sr,Porrati:2010hm}. Therefore, it does not seem possible to use string theory as a ``generator'' of consistent higher-spin theories featuring just a single spin-$s$ field. 

While the problem of higher spin interactions may seem a bit esoteric, the recent applications to the gravitational wave physics have contributed to the Renaissance of the topic, see e.g. \cite{Vines:2017hyw,Arkani-Hamed:2017jhn, Guevara:2018wpp,Chung:2018kqs,Skvortsov:2023jbn,Cangemi:2023bpe}. Indeed, instead of solving Einstein equations for two compact objects one can apply the effective field theory approach to model well-separated compact rotating objects (black holes, neutron stars, etc.) as massive higher-spin particles undergoing specific types of gravitational interactions that cause them to move as if they were in general relativity. Different types of interactions can correspond to different types of compact objects with black holes argued to be described by the simplest theory of this kind. Via the classical double-copy construction one can take the ``square root'' of the problem to search for electromagnetic/non-abelian gauge interactions of massive higher-spin fields instead of the gravitational ones, see e.g. \cite{Monteiro:2014cda,Arkani-Hamed:2019ymq,Guevara:2020xjx,Bern:2010ue}.

In this paper, we reconsider the problem of the massive spin three-half field in a constant electromagnetic background. We prefer to directly analyze the structure of the covariant constraints, which is equivalent to the Stueckelberg approach discussed above. We formulate the most general ansatz for interactions and derive the algebraic system of equations that determines consistent interactions. An explicit solution is obtained as well and, in a sense, our paper is a development of the very important \cite{Porrati:2009bs} that settled some longstanding issues, perhaps, for the first time. Lastly, we perform the transformation to the chiral approach as to reveal the structure of non-minimal couplings needed to restore the parity. 

Additional bits of motivation to study the constant electromagnetic background include: (a) it is inaccessible by the usual amplitude techniques; (b) closed-form expressions for all orders in the electromagnetic field can be obtained, which then can serve as a starting point for the derivative expansion.

The outline of the rest of the paper is as follows. We briefly recall the story of the massive spin three-half field in the Minkowski space. Then, we introduce the minimal gauge interaction and point out where the first obstruction is coming from and how it can be cured. Next, we write down the most general ansatz for electromagnetic and Yang-Mills interactions and analyze the differential consequences of the Lagrangian equations of motion to make sure that the auxiliary fields vanish on-shell. Afterward, we discuss the space of solutions and construct a simple exact one, which is nonpolynomial in the field strength of the background field. Lastly, we perform the transformation to the chiral approach to the leading order in the field strength.

\section{Free massive spin-three-half}
\label{sec:free_field}
Let us start with the $4d$ Rarita-Schwinger \cite{Rarita:1941mf} action already in the spinorial language. The vector-spinor $\psi^\mu$, which is usually considered in the literature, can be decomposed into $(2,1)$, $(1,2)$, $(1,0)$ and $(0,1)$ irreducible representations of the Lorentz algebra $sl(2,\mathbb{C})$:\footnote{$A,B,C,...=1,2$ and $A',B',C',...=1,2$ are the indices of the (anti)-fundamental representations of $sl(2,\mathbb{C})$. They are raised and lowered with the help of $\epsilon^{AB}=-\epsilon^{BA}$ as $v^A=\epsilon^{AB} v_B$, $v_B=v^A\epsilon_{AB}$, idem for primed indices. Note that the rules also apply to $\epsilon^{AB}$ itself and $\epsilon\fdu{A}{B}=-\epsilon\fud{B}{A}=\delta\fdu{A}{B}$. Round brackets denote the symmetrization of the indices enclosed. } $\psi_{ABA'}$ and its conjugate $\bar{\psi}_{AA'B'}$; auxiliary spinor field $\xi_{A}$ and its conjugate $\bar{\xi}_{A'}$, the latter two being the $\gamma$-trace $\gamma_\mu \psi^\mu$ and the former two representing the $\gamma$-trace-free part of $\psi^\mu$. 

Fields $\psi_{ABA'}$ and $\bar{\psi}_{AA'B'}$ are the physical fields for which we need to get the Dirac-like equations. However, there are unphysical longitudinal modes that need to be removed via the transversality constraints 
\besubeqs\label{transcons}
\begin{align}
\partial^{CC'}\psi_{ACC'} &= 0\,,
\label{eq:spin3/2_free_transverse_constraint} \\
\partial^{CC'}\bar{\psi}_{CC'A'} &= 0 \, .
\label{eq:spin3/2_free_transverse_constraint_conj}
\end{align}
\esubeqs
Altogether, there are too many equations for the system to be Lagrangian, and auxiliary fields $\xi^A$ and $\bar{\xi}^{A'}$ help to solve this problem. Indeed, the Lagrangian density of the free field in the Minkowski spacetime is\footnote{Note that the coordinates $x^{AA'}$ and, hence, the derivative $\pl_{AA'}$ are chosen to be anti-Hermitian.}
\begin{align}
\mathcal{L} &= \sqrt{2}\bar{\psi}^{AA'B'}\partial^{C}_{\phantom{C}A'}\psi_{ACB'} + \frac{1}{2}m\Big(\psi^{ABA'}\psi_{ABA'} - \bar{\psi}^{AA'B'}\bar{\psi}_{AA'B'}\Big) \notag \\
&\qquad - 3\sqrt{2}\bar{\xi}^{A'}\partial_{AA'}\xi^{A} + 3m\Big(\xi^{A}\xi_{A} - \bar{\xi}^{A'}\bar{\xi}_{A'}\Big) + \sqrt{2}\Big(\psi^{ABA'}\partial_{AA'}\xi_{B} + \bar{\psi}^{AA'B'}\partial_{AA'}\bar{\xi}_{B'}\Big) \, ,
\label{eq:spin3/2_free_Lagrangian}
\end{align}
where the coefficients are chosen in order to find the desired constraints: the vanishing of the auxiliary fields $\xi^A$, $\bar{\xi}^{A'}$ and the transversality constraint \eqref{transcons}. The equations of motion obtained from this Lagrangian density read
\besubeqs
\begin{align}
{E^{\psi}}_{ABA'} &:= m\psi_{ABA'} + \sqrt{2}\partial_{(A}^{\phantom{(A}B'}\bar{\psi}_{B)A'B'} + \sqrt{2}\partial_{(A|A'|}\xi_{B)} = 0\,,
\label{eq:spin3/2_free_EOM_1} \\[2mm]
{E^{\bar{\psi}}}_{AA'B'} &:= -m\bar{\psi}_{AA'B'} + \sqrt{2}\partial^{C}_{\phantom{C}(A'}\psi_{|AC|B')} + \sqrt{2}\partial_{A(A'}\bar{\xi}_{B')} = 0\,,
\label{eq:spin3/2_free_EOM_2} \\[2mm]
{E^{\xi}}_{A} &:= 6m\xi_{A} - 3\sqrt{2}\partial_{AA'}\bar{\xi}^{A'} - \sqrt{2}\partial^{CC'}\psi_{ACC'} = 0\,,
\label{eq:spin3/2_free_EOM_3} \\[2mm]
{E^{\bar{\xi}}}_{A'} &:= -6m\bar{\xi}_{A'} - 3\sqrt{2}\partial_{AA'}\xi^{A} - \sqrt{2}\partial^{CC'}\bar{\psi}_{CC'A'} = 0 \, .
\label{eq:spin3/2_free_EOM_4}
\end{align}
\esubeqs
The desired constraints can be found by combining the equations of motion and derivatives thereof. For example, the expression
\begin{align}
\partial^{BB'}{E^{\psi}}_{ABB'} + \frac{\sqrt{2}}{2}m{E^{\xi}}_{A} + \frac{1}{2}\partial_{A}^{\phantom{A}A'}{E^{\bar{\xi}}}_{A'} \equiv 3\sqrt{2}m^2\xi_{A}
\label{eq:free_constraint}
\end{align}
gives on-shell the constraint $\xi_A = 0$. 
Equivalently, the following expression
\begin{align}
\partial^{BB'}{E^{\bar{\psi}}}_{BB'A'} - \frac{\sqrt{2}}{2}m{E^{\bar{\xi}}}_{A'} + \frac{1}{2}\partial^{A}_{\phantom{A}A'}{E^{\xi}}_{A} \equiv 3\sqrt{2}m^2\bar{\xi}_{A'}
\end{align}
gives the constraint $\bar{\xi}_{A'} = 0$. By plugging these constraints back into the equations of motion, we obtain the two Dirac-like equations of motion for the physical fields
\besubeqs
\begin{align}
m\psi_{ABA'} + \sqrt{2}\partial_{(A}^{\phantom{(A}B'}\bar{\psi}_{B)A'B'} &= 0\,,
\label{eq:spin3/2_free_Dirac_EOM} \\[2mm]
-m\bar{\psi}_{AA'B'} + \sqrt{2}\partial^{C}_{\phantom{C}(A'}\psi_{|AC|B')} &= 0 \, ,
\label{eq:spin3/2_free_Dirac_EOM_conj}
\end{align}
\esubeqs
from \eqref{eq:spin3/2_free_EOM_1} and \eqref{eq:spin3/2_free_EOM_2}, and the transversality constraints \eqref{transcons} from \eqref{eq:spin3/2_free_EOM_3}, \eqref{eq:spin3/2_free_EOM_4}. The relative coefficient between the kinetic and the mass terms is chosen to recover the familiar Klein-Gordon equation of motion
\begin{align}
\big(\Box - m^2\big)\psi_{ABA'} = 0 \, ,
\end{align}
where $\Box := \partial_{AA'}\partial^{AA'}$, which is obtained by solving \eqref{eq:spin3/2_free_Dirac_EOM_conj} with respect to $\bar{\psi}_{AA'B'}$ and plugging it into  \eqref{eq:spin3/2_free_Dirac_EOM}.

\section{Minimal Electromagnetic/Yang-Mills interactions}
\label{sec:}
In this section, we attempt to introduce the minimal electromagnetic/Yang-Mills interactions and show how they modify/destroy the constraints, the problem that can be cured by introducing higher order/nonminimal interactions. The covariant derivative is defined as
\begin{align}
    D&= d +\mathcal{A}\,, && D= dx^\mu\, e^{AA'}_\mu D_{AA'}\,,
\end{align}
where $\mathcal{A}\equiv \mathcal{A}_\mu\, dx^\mu$ is the electromagnetic/Yang-Mills gauge field. The vierbein $e^{AA'}_\mu$ is a bit of an exaggeration since we consider the Minkowski spacetime. Given that $u(N) \subset so(2N)$ we consider $so(2N)$ gauging, i.e. the fields are in the vector representation of $so(2N)$, e.g. $\phi\equiv \phi^i$, $i,j,k,...=1,...,2N$. The gauge field $\mathcal{A}$ in the adjoint is $\mathcal{A}^{ij}=-\mathcal{A}^{ji}$. Whenever no ambiguity arises we omit the $so(2N)$-indices. The commutator of two covariant derivatives  
\begin{align}
[{D}_{AA'},{D}_{BB'}]\bullet := {F}_{ABA'B'}\bullet \equiv \tfrac{1}{2}\epsilon_{A'B'}{F}_{AB}\bullet + \tfrac{1}{2}\epsilon_{AB}{F}_{A'B'}\bullet \, ,
\end{align}
defines the field strength. Whenever we write $F$ we mean the full field strength ${F}_{ABA'B'}$, i.e. both its selfdual $F_{AB}$ and anti-selfdual $F_{A'B'}$ components, e.g. $g(F)$ means a function $g(F_{AB},{F}_{A'B'})$. Whenever two indices of the covariant derivatives are contracted we find
\besubeqs
\begin{align}
{D}_{AA'}{D}_{B}^{\phantom{B}A'}\bullet &\equiv \tfrac{1}{2}[{D}_{AA'},{D}_{B}^{\phantom{B}A'}]\bullet + \tfrac{1}{2}\{{D}_{AA'},{D}_{B}^{\phantom{B}A'}\}\bullet \equiv \tfrac{1}{2}{F}_{AB}\bullet + \tfrac{1}{2}\epsilon_{AB}\square\bullet \,, \\
{D}_{AA'}{D}_{\phantom{A}B'}^{A}\bullet &\equiv \tfrac{1}{2}[{D}_{AA'},{D}_{\phantom{A}B'}^{A}]\bullet + \tfrac{1}{2}\{{D}_{AA'},{D}_{\phantom{A}B'}^{A}\}\bullet \equiv \tfrac{1}{2}{F}_{A'B'}\bullet + \tfrac{1}{2}\epsilon_{A'B'}\square\bullet \, ,
\end{align}
\esubeqs
Now we simply replace all partial derivatives with the covariant ones in Lagrangian \eqref{eq:spin3/2_free_Lagrangian}, which gives 
\begin{align}
\mathcal{L} &= \sqrt{2}\bar{\psi}^{AA'B'}D^{C}_{\phantom{C}A'}\psi_{ACB'} + \frac{1}{2}m\Big(\psi^{ABA'}\psi_{ABA'} - \bar{\psi}^{AA'B'}\bar{\psi}_{AA'B'}\Big) \notag \\
&\qquad - 3\sqrt{2}\bar{\xi}^{A'}D_{AA'}\xi^{A} + 3m\Big(\xi^{A}\xi_{A} - \bar{\xi}^{A'}\bar{\xi}_{A'}\Big) + \sqrt{2}\Big(\psi^{ABA'}D_{AA'}\xi_{B} + \bar{\psi}^{AA'B'}D_{AA'}\bar{\xi}_{B'}\Big) \, .
\label{eq:Lagrangian_minimal_int}
\end{align}
The equations of motion change accordingly
\besubeqs
\begin{align}
{E^{\psi}}_{ABA'} &:= m\psi_{ABA'} + \sqrt{2}D_{(A}^{\phantom{(A}B'}\bar{\psi}_{B)A'B'} + \sqrt{2}D_{(A|A'|}\xi_{B)} = 0\,,
\label{eq:minimal_int_EOM_1} \\[2mm]
{E^{\bar{\psi}}}_{AA'B'} &:= -m\bar{\psi}_{AA'B'} + \sqrt{2}D^{C}_{\phantom{C}(A'}\psi_{|AC|B')} + \sqrt{2}D_{A(A'}\bar{\xi}_{B')} = 0\,,
\label{eq:minimal_int__EOM_2} \\[2mm]
{E^{\xi}}_{A} &:= 6m\xi_{A} - 3\sqrt{2}D_{AA'}\bar{\xi}^{A'} - \sqrt{2}D^{CC'}\psi_{ACC'} = 0\,,
\label{eq:minimal_int__EOM_3} \\[2mm]
{E^{\bar{\xi}}}_{A'} &:= -6m\bar{\xi}_{A'} - 3\sqrt{2}D_{AA'}\xi^{A} - \sqrt{2}D^{CC'}\bar{\psi}_{CC'A'} = 0 \, .
\label{eq:minimal_int__EOM_4}
\end{align}
\esubeqs
The constraint in the case of the minimal interaction must have the same form \eqref{eq:free_constraint} but with covariant derivatives instead of partial ones
\begin{align}
D^{BB'}{E^{\psi}}_{ABB'} + \frac{\sqrt{2}}{2}m{E^{\xi}}_{A} + \frac{1}{2}D_{A}^{\phantom{A}A'}{E^{\bar{\xi}}}_{A'} = 0\, .
\label{eq:minimal_int_constraint}
\end{align}
It reduces to
\begin{align}
3\sqrt{2}m^2\xi_A + \sqrt{2}F_{AB}\xi^B - \frac{\sqrt{2}}{2}F^{B'C'}\bar{\psi}_{AB'C'} = 0 \, .
\label{eq:minimal_int_constraint_developed}
\end{align}
The first two terms one can rewrite as $M\fdu{A}{B}\xi_B$ where $M\fdu{A}{B}$ is close to the unit matrix up to $3\sqrt{2}m^2$ since $F$ is assumed small. It is the last term that prevents us from getting $\xi^A=0$. Likewise, we do not recover the transversality constraints.\footnote{Maybe a more detailed analysis can still prove the equations be admissible at least in the sense of describing the right number of degrees of freedom. Indeed, the last expression seems consistent with the analysis of \cite{Zinoviev:2009hu} based on the Stueckelberg gauge symmetry (to recover the constraint one needs to get the equation for the Stuckelberg field and set it to zero.} 

\section{Slightly nonminimal Yang-Mills interactions}
\label{sec:first_order_int}
As is well-known, the problem found above can partially be solved by adding nonminimal interactions, i.e. interactions that have $F$. Since the unwanted term in the constraint \eqref{eq:minimal_int_constraint_developed} is of the first order in $F$, one expects that it can be canceled by adding nonminimal terms linear in $F$ into the Lagrangian \eqref{eq:Lagrangian_minimal_int} and also, importantly, into the constraint \eqref{eq:minimal_int_constraint}. For the Lagrangian we can write
\begin{align}
\mathcal{L} = &\sqrt{2}\bar{\psi}^{AA'B'}D^{C}_{\phantom{C}A'}\psi_{ACB'} + \frac{1}{2}m\Big(\psi^{ABA'}\psi_{ABA'} - \bar{\psi}^{AA'B'}\bar{\psi}_{AA'B'}\Big) \notag \\
&- 3\sqrt{2}\bar{\xi}^{A'}D_{AA'}\xi^{A} + 3m\Big(\xi^{A}\xi_{A} - \bar{\xi}^{A'}\bar{\xi}_{A'}\Big) + \sqrt{2}\Big(\psi^{ABA'}D_{AA'}\xi_{B} + \bar{\psi}^{AA'B'}D_{AA'}\bar{\xi}_{B'}\Big) \notag \\
&+ b_1\Big(\xi_AF^{AB}\xi_B - \bar{\xi}_{A'}F^{A'B'}\bar{\xi}_{B'}\Big) + b_2\Big(\psi_{ABA'}F^A_{\phantom{A}C}\psi^{BCA'} - \bar{\psi}_{AA'B'}F^{A'}_{\phantom{A'}C'}\bar{\psi}^{AB'C'}\Big) \notag \\
&+ b_3\Big(\psi_{ABA'}F^{A'}_{\phantom{A'}B'}\psi^{ABB'} - \bar{\psi}_{AA'B'}F^{A}_{\phantom{A}B}\bar{\psi}^{BA'B'}\Big) + b_4\Big(\psi_{ABA'}F^{AB}\bar{\xi}^{A'} - \bar{\psi}_{AA'B'}F^{A'B'}\xi^A\Big) \, .
\label{eq:Lagrangian_1st_order_int}
\end{align}
It gives the following equations of motion
\besubeqs
\begin{align}
{E^{\psi}}_{ABA'} &:= m\psi_{ABA'} + \sqrt{2}D_{(A}^{\phantom{(A}B'}\bar{\psi}_{B)A'B'} + \sqrt{2}D_{(A|A'|}\xi_{B)} \notag \\
&\qquad + 2b_2F_{(A}^{\phantom{(A}C}\psi_{B)CA'} + 2b_3F_{A'}^{\phantom{A'}B'}\psi_{ABB'} + b_4F_{AB}\bar{\xi}_{A'} = 0 \, ,
\label{eq:1st_order_int_EOM_1} \\[2mm]
{E^{\bar{\psi}}}_{AA'B'} &:= -m\bar{\psi}_{AA'B'} + \sqrt{2}D^{C}_{\phantom{C}(A'}\psi_{|AC|B')} + \sqrt{2}D_{A(A'}\bar{\xi}_{B')} \notag \\
&\qquad - 2b_2F_{(A'|}^{\phantom{(A'|}C'}\bar{\psi}_{A|B')C'} - 2b_3F_{A}^{\phantom{A}B}\bar{\psi}_{BA'B'} - b_4F_{A'B'}\xi_{A} = 0 \, ,
\label{eq:1st_order_int__EOM_2} \\[2mm]
{E^{\xi}}_{A} &:= 6m\xi_{A} - 3\sqrt{2}D_{AA'}\bar{\xi}^{A'} - \sqrt{2}D^{CC'}\psi_{ACC'} + 2b_1F_{AB}\xi^{B} + b_4F^{A'B'}\bar{\psi}_{AA'B'} = 0 \, ,
\label{eq:1st_order_int__EOM_3} \\[2mm]
{E^{\bar{\xi}}}_{A'} &:= -6m\bar{\xi}_{A'} - 3\sqrt{2}D_{AA'}\xi^{A} - \sqrt{2}D^{CC'}\bar{\psi}_{CC'A'} - 2b_1F_{A'B'}\bar{\xi}^{B'} - b_4F^{AB}\psi_{ABA'} = 0 \, .
\label{eq:1st_order_int__EOM_4}
\end{align}
\esubeqs
We also do not forget to add to the constraint all possible terms linear in $F$, which gives
\begin{align}
D^{BB'}{E^{\psi}}_{ABB'} + \frac{\sqrt{2}}{2}m{E^{\xi}}_{A} + \frac{1}{2}D_{A}^{\phantom{A}A'}{E^{\bar{\xi}}}_{A'} + c_1F_A^{\phantom{A}B}{E^{\xi}}_B + c_2F^{A'B'}{E^{\bar{\psi}}}_{AA'B'} = 0\, .
\label{eq:1st_order_int_constraint}
\end{align}
In order to obtain the desired constraint, the vanishing of the auxiliary field, we need to cancel all terms with derivatives, which leads to
\begin{align}
b_1 = -\frac{1}{m} \,, \quad \quad b_2 = 0 \,, \quad \quad b_3 = -\frac{1}{2m} \,, \quad \quad b_4 = 0 \,, \quad \quad c_1 = 0 \,, \quad \quad c_2 = -\frac{\sqrt{2}}{2m} \, .
\label{eq:1st_order_int_coeff}
\end{align}
The constraint reduces to
\begin{align}
3\sqrt{2}m^2\xi_A - \frac{1}{m}D^{BB'}F_{B'}^{\phantom{B'}C'}\psi_{ABC'} + \frac{1}{m}D_A^{\phantom{A}A'}F_{A'B'}\bar{\xi}^{B'} - \frac{\sqrt{2}}{2m^2}F^{A'B'}F_A^{\phantom{A}B}\bar{\psi}_{BA'B'} = 0 \, .
\end{align}
We assume that the background gauge field satisfies its equations of motion, i.e. $D_A^{\phantom{A}B'}F_{A'B'} = 0$, $D^{B}_{\phantom{B}A'}F_{AB} = 0$ (one can add a source as well), which eliminates the 2nd and the 3rd terms. As a result, we are left with, cf. \cite{Porrati:2009bs},
\begin{align}
3\sqrt{2}m^2\xi_A - \frac{\sqrt{2}}{2m^2}F^{A'B'}F_A^{\phantom{A}B}\bar{\psi}_{BA'B'} = 0 \, .
\label{eq:1st_order_int_constraint_develop}
\end{align}
We managed to get rid of the $F\psi$-term, but are left with the $F^2\psi$ one. If the field strength is small enough, this term can effectively be set to zero and we recover $\xi_A = 0$. By using this in the third and fourth equations of motion, we obtain the (covariant) transversality constraints $D^{CC'}\psi_{ACC'} = 0$, $D^{CC'}\psi_{CC'A'} = 0$. Therefore, the Lagrangian describes the right number of degrees of freedom if the Yang-Mills field is small enough. The Lagrangian density reads
\begin{align}
\mathcal{L} = &\sqrt{2}\bar{\psi}^{AA'B'}D^{C}_{\phantom{C}A'}\psi_{ACB'} + \frac{1}{2}m\Big(\psi^{ABA'}\psi_{ABA'} - \bar{\psi}^{AA'B'}\bar{\psi}_{AA'B'}\Big) \notag \\
&- 3\sqrt{2}\bar{\xi}^{A'}D_{AA'}\xi^{A} + 3m\Big(\xi^{A}\xi_{A} - \bar{\xi}^{A'}\bar{\xi}_{A'}\Big) + \sqrt{2}\Big(\psi^{ABA'}D_{AA'}\xi_{B} + \bar{\psi}^{AA'B'}D_{AA'}\bar{\xi}_{B'}\Big) \notag \\
&- \frac{1}{m}\Big(\xi_AF^{AB}\xi_B - \bar{\xi}_{A'}F^{A'B'}\bar{\xi}_{B'}\Big) - \frac{1}{2m}\Big(\psi_{ABA'}F^{A'}_{\phantom{A'}B'}\psi^{ABB'} - \bar{\psi}_{AA'B'}F^{A}_{\phantom{A}B}\bar{\psi}^{BA'B'}\Big) + \mathcal{O}(F^2) \, ,
\label{eq:Lagrangian_1st_order_int_with_valued_coeff}
\end{align}
as obtained from \eqref{eq:Lagrangian_1st_order_int} with \eqref{eq:1st_order_int_coeff}. 

\section{Consistent Yang-Mills interactions}
We found a consistent Lagrangian for massive spin three-half fields interacting with a small vacuum Yang-Mills field. While for some practical applications the Lagrangian may suffice, it is interesting to solve the problem without making any truncations. After introducing the minimal interaction, the undesired term in the constraint is of the form $F\bar{\psi}$, see \eqref{eq:minimal_int_constraint_developed}. By trying to cancel it with the first order nonminimal terms, the undesired term in the constraint was pushed to $\bar{F}F\bar{\psi}$, see \eqref{eq:1st_order_int_constraint_develop}. It is clear that in trying to cancel an $F^n$-order undesired terms in the constraint by introducing the next order nonminimal terms in the Lagrangian should give some $F^{n+1}$-order undesired terms in the new constraint. Therefore, let us construct a Lagrangian density with the most general nonminimal interactions, which are parameterized by a number of functions of $F$. It reads
\begin{align}
\mathcal{L} = &\sqrt{2}\bar{\psi}^{AA'B'}D^{C}_{\phantom{C}A'}\psi_{ACB'} + \frac{1}{2}m\Big(\psi^{ABA'}\psi_{ABA'} - \bar{\psi}^{AA'B'}\bar{\psi}_{AA'B'}\Big) \notag \\
&- 3\sqrt{2}\bar{\xi}^{A'}D_{AA'}\xi^{A} + 3m\Big(\xi^{A}\xi_{A} - \bar{\xi}^{A'}\bar{\xi}_{A'}\Big) + \sqrt{2}\Big(\psi^{ABA'}D_{AA'}\xi_{B} + \bar{\psi}^{AA'B'}D_{AA'}\bar{\xi}_{B'}\Big) \notag \\
&+ \Big(\psi^{ABA'}g_1(F)_{AB|CD;A'|B'}\psi^{CDB'} - \bar{\psi}^{AA'B'}\overline{g_1(F)}_{A|B;A'B'|C'D'}\bar{\psi}^{BC'D'}\Big) \notag \\
&+ \Big(\xi^Ag_2(F)_{A|B}\xi^B - \bar{\xi}^{A'}\overline{g_2(F)}_{A'|B'}\bar{\xi}^{B'}\Big) \notag \\
&+ \Big(\psi^{ABA'}g_3(F)_{AB;A'|B'}\bar{\xi}^{B'} - \bar{\psi}^{AA'B'}\overline{g_3(F)}_{A|B;A'B'}\xi^B\Big) \, ,
\label{eq:Lagrangian_general_YM_int}
\end{align}
where $g_1(F)_{AB|CD;A'|B'}$, $g_2(F)_{A|B}$ and $g_3(F)_{AB;A'|B'}$ are arbitrary functions of $F$ which vanish at $F=0$. Their index structure is the most general taking into account the index structure of the fields with which they are contracted. The notation above means that the indices that are not separated by ``$|$'' are symmetrized, and ``$;$'' separates primed and unprimed indices. Note that with $F_{AB}$, $F_{A'B'}$ and $\epsilon_{AB}$, $\epsilon_{A'B'}$ we can only construct tensors with an even number of indices of each sort. Therefore, we cannot add something like $g_{AB|C;A'}(F)\psi^{ABA'} \xi^C$ to the action unless derivatives of $F$ are introduced. Let us define two more functions of this kind in order to generalize the constraint \eqref{eq:minimal_int_constraint}
\begin{align}
D^{BB'}{E^{\psi}}_{ABB'} + \frac{\sqrt{2}}{2}m{E^{\xi}}_{A} + \frac{1}{2}D_{A}^{\phantom{A}A'}{E^{\bar{\xi}}}_{A'} + h_1(F)_{A|B;A'B'}{E^{\bar{\psi}}}^{BA'B'} + h_2(F)_{A|B}{E^{\xi}}^B = 0\, .
\label{eq:general_YM_int_constraint}
\end{align}
In order to simplify the problem let us restrict to the constant background, i.e. $DF=0$. Also, functions $g_{1,2,3}$ and $h_{1,2}$ are not irreducible tensors yet. It is usually a good idea to decompose everything into irreducible tensors, which gives
{\allowdisplaybreaks
\besubeqs
\begin{align}
g_1(F)_{AB|CD;A'|B'} &= g_1(F)_{ABCDA'B'} + \frac{1}{2}g_1(F)_{ABCD}\epsilon_{A'B'} \notag \\
&\quad + g_1(F)_{(A|\textcolor{red}{(}C\textcolor{red}{|}A'B'}\epsilon_{|B)\textcolor{red}{|}D\textcolor{red}{)}} + \frac{1}{2}g_1(F)_{(A|\textcolor{red}{(}C\textcolor{red}{|}}\epsilon_{|B)\textcolor{red}{|}D\textcolor{red}{)}}\epsilon_{A'B'} \notag \\
&\quad + \frac{1}{3}g_1(F)_{A'B'}\epsilon_{(A|C}\epsilon_{|B)D} + \frac{1}{6}g_1(F)\epsilon_{(A|C}\epsilon_{|B)D}\epsilon_{A'B'} \,,
\label{eq:decomp_g1} \\[3mm]
g_2(F)_{A|B} &= g_2(F)_{AB} + \frac{1}{2}g_2(F)\epsilon_{AB} \, ,
\label{eq:decomp_g2} \\[3mm]
g_3(F)_{AB;A'|B'} &= g_3(F)_{ABA'B'} + \frac{1}{2}g_3(F)_{AB}\epsilon_{A'B'} \, ,
\label{eq:decomp_g3} \\[3mm]
h_1(F)_{A|B;A'B'} &= h_1(F)_{ABA'B'} + \frac{1}{2}h_1(F)_{A'B'}\epsilon_{AB} \, ,
\label{eq:decomp_h1} \\[3mm]
h_2(F)_{A|B} &= h_2(F)_{AB} + \frac{1}{2}h_2(F)\epsilon_{AB} \, ,
\label{eq:decomp_h2}
\end{align}
\esubeqs}

\noindent where all new functions on the right are completely symmetric in their indices. Let us also recall that all the fermions are in the vector representation of $so(2N)$ and, hence, all the functions have two $so(2N)$ indices, all of which are buried now in our notation. Since it may lead to some confusion when getting the equations of motion let us give some examples, e.g.
\begin{align}\notag
\psi^{ABA'}g_1(F)_{AB|CD;A'|B'}\psi^{CDB'} \equiv \psi^{ABA'}_ig_1(F)^{ij}_{AB|CD;A'|B'}\psi^{CDB'}_j = -\psi^{ABA'}_ig_1(F)^{ji}_{CD|AB;B'|A'}\psi^{CDB'}_j \, .
\end{align}
Here we performed the standard manipulations by swapping the two fermions and renaming the dummy indices. This leads to the following property of the function $g_1$
\begin{align}
g_1(F)^{ij}_{AB|CD;A'|B'} = -g_1(F)^{ji}_{CD|AB;B'|A'} \, .
\label{eq:symm_prop_g_1}
\end{align}
Similarly, the function $g_2$ has the following property 
\begin{align}
g_2(F)_{A|B}^{ij} = -g_2(F)_{B|A}^{ji} \, .
\label{eq:symm_prop_g_2}
\end{align}
These symmetry properties allow one to calculate the following contributions to the equations of motion
\besubeqs
\begin{align}
&\frac{\delta}{\delta\psi^{ABA'}}\Big(\psi^{EFE'}g_1(F)_{EF|CD;E'|B'}\psi^{CDB'}\Big) = 2g_1(F)_{AB|CD;A'|B'}\psi^{CDB'} \, , \\
&\frac{\delta}{\delta\xi^A}\Big(\xi^Cg_2(F)_{C|B}\xi^B\Big) = 2g_2(F)_{A|B}\xi^B \, .
\end{align}
\esubeqs
The function $g_3$ couples $\psi$ and $\bar{\xi}$ and does not have any additional symmetry properties. The corresponding contributions to the equations of motion read
\begin{align}
\frac{\delta}{\delta\psi^{ABA'}}\Big(\psi^{CDC'}g_3(F)_{CD;C'|B'}\bar{\xi}^{B'}\Big) = g_3(F)_{AB;A'|B'}\bar{\xi}^{B'} \, ,
\end{align}
for $\psi^{ABA'}$ and
\begin{align}
\frac{\delta}{\delta\bar{\xi}^{A'}}\Big(\psi^{ABC'}g_3(F)_{AB;C'|B'}\bar{\xi}^{B'}\Big) &\equiv \frac{\delta}{\delta\bar{\xi}^{A'}_i}\Big(\psi^{ABC'}_jg_3(F)^{jk}_{AB;C'|B'}\bar{\xi}_k^{B'}\Big) = -g_3(F)^{ji}_{AB;C'|A'}\psi^{ABC'}_j \, ,
\end{align}
for $\bar{\xi}$. In order to enjoy the index-free notation again, we need to invoke the transposed of the matrix $g_3$ with respect to $so(2N)$ indices, denoted $g^T_3$,
\begin{align}
g^T_3(F)^{ij}_{AB;A'|B'} := g_3(F)^{ji}_{AB;A'|B'} \, .
\end{align}
The last term in the equations of motion now becomes
\begin{align}
\frac{\delta}{\delta\bar{\xi}^{A'}}\Big(\psi^{ABC'}g_3(F)_{AB;C'|B'}\bar{\xi}^{B'}\Big) = -g^T_3(F)_{AB;B'|A'}\psi^{ABB'}\, .
\end{align}
Finally, the equations of motion are
\besubeqs
\begin{align}
{E^{\psi}}_{ABA'} &:= m\psi_{ABA'} + \sqrt{2}D_{(A}^{\phantom{(A}B'}\bar{\psi}_{B)A'B'} + \sqrt{2}D_{(A|A'|}\xi_{B)} + 2g_1(F)_{ABCDA'B'}\psi^{CDB'} \notag \\
&\quad\; - g_1(F)_{ABCD}\psi^{CD}_{\phantom{CD}A'} - 2g_1(F)_{(A|CA'B'}\psi_{|B)}^{\phantom{|B)}CB'} + g_1(F)_{(A|C}\psi_{|B)\phantom{C}A'}^{\phantom{|B)}C} \notag \\
&\quad\; + \frac{2}{3}g_1(F)_{A'B'}\psi_{AB}^{\phantom{AB}B'} - \frac{1}{3}g_1(F)\psi_{ABA'} + g_3(F)_{ABA'B'}\bar{\xi}^{B'} - \frac{1}{2}g_3(F)_{AB}\bar{\xi}_{A'} = 0 \, ,
\label{eq:general_int_EOM_1} \\[3mm]
{E^{\bar{\psi}}}_{AA'B'} &:= -m\bar{\psi}_{AA'B'} + \sqrt{2}D^{C}_{\phantom{C}(A'}\psi_{|AC|B')} + \sqrt{2}D_{A(A'}\bar{\xi}_{B')} - 2\overline{g_1(F)}_{ABA'B'C'D'}\bar{\psi}^{BC'D'} \notag \\
&\quad\; + \overline{g_1(F)}_{A'B'C'D'}\bar{\psi}^{\phantom{A}C'D'}_{A} + 2\overline{g_1(F)}_{ABC'(A'}\bar{\psi}_{\phantom{BC'}B')}^{BC'} - \overline{g_1(F)}_{(A'|C'}\bar{\psi}_{A|B')}^{\phantom{A|B')}C'} \notag \\
&\quad\; - \frac{2}{3}\overline{g_1(F)}_{AB}\bar{\psi}_{\phantom{B}A'B'}^{B} + \frac{1}{3}\overline{g_1(F)}\bar{\psi}_{AA'B'} - \overline{g_3(F)}_{ABA'B'}\xi^{B} + \frac{1}{2}\overline{g_3(F)}_{A'B'}\xi_{A} = 0 \, ,
\label{eq:general_int_EOM_2} \\[3mm]
{E^{\xi}}_{A} &:= 6m\xi_{A} - 3\sqrt{2}D_{AA'}\bar{\xi}^{A'} - \sqrt{2}D^{CC'}\psi_{ACC'} + 2g_2(F)_{AB}\xi^{B} \notag \\
&\quad\; - g_2(F)\xi_{A} + \overline{g^T_3(F)}_{ABA'B'}\bar{\psi}^{BA'B'} + \frac{1}{2}\overline{g^T_3(F)}_{A'B'}\bar{\psi}_A^{\phantom{A}A'B'} = 0 \, ,
\label{eq:general_int_EOM_3} \\[3mm]
{E^{\bar{\xi}}}_{A'} &:= -6m\bar{\xi}_{A'} - 3\sqrt{2}D_{AA'}\xi^{A} - \sqrt{2}D^{CC'}\bar{\psi}_{CC'A'} - 2\overline{g_2(F)}_{A'B'}\bar{\xi}^{B'} \notag \\
&\quad\; + \overline{g_2(F)}\bar{\xi}_{A'} - g^T_3(F)_{ABA'B'}\psi^{ABB'} - \frac{1}{2}g^T_3(F)_{AB}\psi^{AB}_{\phantom{AB}A'} = 0 \, .
\label{eq:general_int_EOM_4}
\end{align}
\esubeqs
The constraint \eqref{eq:general_YM_int_constraint} can be unfolded into
\begin{align}
D^{BB'}{E^{\psi}}_{ABB'} &+ \frac{\sqrt{2}}{2}m{E^{\xi}}_{A} + \frac{1}{2}D_{A}^{\phantom{A}A'}{E^{\bar{\xi}}}_{A'} + h_1(F)_{ABA'B'}{E^{\bar{\psi}}}^{BA'B'} \notag \\
&- \frac{1}{2}h_1(F)_{A'B'}{E^{\bar{\psi}}}_{A}^{\phantom{A}A'B'} + h_2(F)_{AB}{E^{\xi}}^B - \frac{1}{2}h_2(F){E^{\xi}}_A = 0\, .
\label{eq:general_YM_int_constraint_decomposed}
\end{align}
Now, we can develop the constraint by using the expressions of the equations of motion, which leads to a lengthy expression in Appendix \ref{app:bigmess}. By setting to zero all coefficients in front of $D\xi$, $\psi$ and $D\psi$ terms we get the following system of linear equations
\besubeqs
\begin{center}
\begin{tabular}{p{0.46\linewidth}p{0.5\linewidth}}
{\begin{align}
g_1(F)_{ABCDA'B'} &= 0 \,,
\label{eq:eq_function_g_1(4,2)} \\[2mm]
g_1(F)_{ABCD} &= 0\,,
\label{eq:eq_function_g_1(4,0)} \\[2mm]
g_1(F)_{ABA'B'} &= -\frac{1}{2}g_3(F)_{ABA'B'}\,,
\label{eq:eq_function_g_1(2,2)} \\[2mm]
g_1(F)_{AB} &= 0\,,
\label{eq:eq_function_g_1(2,0)} \\[2mm]
g_1(F)_{A'B'} &= -\frac{3}{2}\overline{g_2(F)}_{A'B'}\,,
\label{eq:eq_function_g_1(0,2)} \\[2mm]
g_1(F) &= \frac{1}{2}\overline{g_2(F)}\,,
\label{eq:eq_function_g_1(0,0)}
\end{align}} & {\begin{align}
g^T_3(F)_{ABA'B'} &= g_3(F)_{ABA'B'}\,,
\label{eq:eq_function_g^T_3(2,2)} \\[2mm]
g_3(F)_{AB} &= 0 = g^T_3(F)_{AB}\,,
\label{eq:eq_function_g^T_3(2,0)} \\[2mm]
h_1(F)_{ABA'B'} &= -\frac{\sqrt{2}}{2}g_3(F)_{ABA'B'}\,,
\label{eq:eq_function_h_1(2,2)} \\[2mm]
h_1(F)_{A'B'} &= -\sqrt{2}\ \overline{g_2(F)}_{A'B'}\,,
\label{eq:eq_function_h_1(0,2)} \\[2mm]
h_2(F)_{AB} &= 0\,,
\label{eq:eq_function_h_2(2,0)} \\[2mm]
h_2(F) &= \frac{\sqrt{2}}{6}\overline{g_2(F)}\,,
\label{eq:eq_function_h_2(0,0)}
\end{align}}
\end{tabular}
\end{center}
and two quadratic equations
\begin{align}
F_{A'B'} + m\overline{g_2(F)}_{A'B'} - &\frac{1}{6}\overline{g_2(F)}_{A'B'}g_2(F) +\notag\\ &- \frac{1}{2}g_3(F)_{ABA'B'}g_2(F)^{AB} +  \frac{1}{2}g_3(F)_{ABC'(A'}\overline{g_3(F)}^{ABC'}_{\phantom{ABC'}B')} = 0\,,
\label{eq:eq_g_2_g_3_(0,2)}
\end{align}
\begin{align}
&m\Big(g_3(F)_{ABA'B'} + \overline{g_3(F)}_{ABA'B'}\Big) - \frac{1}{6}\Big(g_3(F)_{ABA'B'}g_2(F) + \overline{g_2(F)}\ \overline{g_3(F)}_{ABA'B'}\Big) \notag \\
&- \Big(g_3(F)_{(A|CA'B'}g_2(F)^{C}_{\phantom{C}|B)} + \overline{g_2(F)}_{(B'|C'}\overline{g_3(F)}_{AB|A')}^{\phantom{AB|A')}C'}\Big) \notag \\
& + \overline{g_2(F)}_{A'B'}g_2(F)_{AB} + g_3(F)_{(A|CC'(A'|}\overline{g_3(F)}_{|B)\phantom{C}|B')}^{\phantom{|B)}C\phantom{|A')}C'} = 0\,.
\label{eq:eq_g_2_g_3_(2,2)}
\end{align}
\esubeqs
We have two nonlinear algebraic equations that constrain functions $g_2(F)_{AB}$, $g_2(F)$ and $g_3(F)_{ABA'B'}$. Then, all the other functions, $g_1$'s, $h_1$'s, $h_2$ and even $g^T_3$, are determined by $g_2(F)_{AB}$, $g_2(F)$ and $g_3(F)_{ABA'B'}$. In particular, note that the relation \eqref{eq:eq_function_g^T_3(2,2)} implies that $g_3(F)_{ABA'B'}$ is a symmetric matrix. Since all these equations have to be satisfied, the equations of motion \eqref{eq:general_int_EOM_1}, \eqref{eq:general_int_EOM_2}, \eqref{eq:general_int_EOM_3} and \eqref{eq:general_int_EOM_4} become
\begin{align}
{E^{\psi}}_{ABA'} &:= m\psi_{ABA'} + \sqrt{2}D_{(A}^{\phantom{(A}B'}\bar{\psi}_{B)A'B'} + \sqrt{2}D_{(A|A'|}\xi_{B)} + g_3(F)_{(A|CA'B'}\psi_{|B)}^{\phantom{|B)}CB'} \notag \\
&\quad\; - \overline{g_2(F)}_{A'B'}\psi_{AB}^{\phantom{AB}B'} - \frac{1}{6}\overline{g_2(F)}\psi_{ABA'} + g_3(F)_{ABA'B'}\bar{\xi}^{B'} = 0 \, ,
\label{eq:general_int_EOM_1_valued} \\[3mm]
{E^{\bar{\psi}}}_{AA'B'} &:= -m\bar{\psi}_{AA'B'} + \sqrt{2}D^{C}_{\phantom{C}(A'}\psi_{|AC|B')} + \sqrt{2}D_{A(A'}\bar{\xi}_{B')} - \overline{g_3(F)}_{ABC'(A'}\bar{\psi}_{\phantom{BC'}B')}^{BC'} \notag \\
&\quad\; + g_2(F)_{AB}\bar{\psi}_{\phantom{B}A'B'}^{B} + \frac{1}{6}g_2(F)\bar{\psi}_{AA'B'} - \overline{g_3(F)}_{ABA'B'}\xi^{B} = 0 \, ,
\label{eq:general_int_EOM_2_valued} \\[3mm]
{E^{\xi}}_{A} &:= 6m\xi_{A} - 3\sqrt{2}D_{AA'}\bar{\xi}^{A'} - \sqrt{2}D^{CC'}\psi_{ACC'} +\notag \\&  \qquad \qquad +2g_2(F)_{AB}\xi^{B} - g_2(F)\xi_{A} + \overline{g_3(F)}_{ABA'B'}\bar{\psi}^{BA'B'} = 0 \, ,
\label{eq:general_int_EOM_3_valued} \\[3mm]
{E^{\bar{\xi}}}_{A'} &:= -6m\bar{\xi}_{A'} - 3\sqrt{2}D_{AA'}\xi^{A} - \sqrt{2}D^{CC'}\bar{\psi}_{CC'A'} - 2\overline{g_2(F)}_{A'B'}\bar{\xi}^{B'} + \notag\\&\qquad \qquad +\overline{g_2(F)}\bar{\xi}_{A'} - g_3(F)_{ABA'B'}\psi^{ABB'} = 0 \, ,
\label{eq:general_int_EOM_4_valued}
\end{align}
The constraint \eqref{eq:general_YM_int_constraint_decomposed} reduces to
\begin{align}
D^{BB'}{E^{\psi}}_{ABB'} + \frac{\sqrt{2}}{2}m{E^{\xi}}_{A} + \frac{1}{2}D_{A}^{\phantom{A}A'}{E^{\bar{\xi}}}_{A'} &-\frac{\sqrt{2}}{2}g_3(F)_{ABA'B'}{E^{\bar{\psi}}}^{BA'B'} \notag \\
&+ \frac{\sqrt{2}}{2}\overline{g_2(F)}_{A'B'}{E^{\bar{\psi}}}_{A}^{\phantom{A}A'B'} - \frac{\sqrt{2}}{12}\overline{g_2(F)}{E^{\xi}}_{A} = 0\, .
\label{eq:general_YM_int_constraint_valued}
\end{align}
Plugging in the equations of motion and taking into account the constraints for $g$'s gives 
\begin{align}
\Bigg(\Big(& F_{AB} + mg_2(F)_{AB} - \frac{1}{6}\overline{g_2(F)}g_2(F)_{AB} - \frac{1}{2}\overline{g_2(F)}_{A'B'}\overline{g_3(F)}_{AB}^{\phantom{AB}A'B'} + \frac{1}{2}g_3(F)_{(A|CA'B'}\overline{g_3(F)}_{|B)}^{\phantom{|B)}CA'B'}\Big) \notag \\
+& \Big( - 3m^2 + \frac{1}{2}mg_2(F) - \frac{1}{2}m\overline{g_2(F)} + \frac{1}{12}\overline{g_2(F)}g_2(F) + \frac{1}{4}g_3(F)_{CDA'B'}\overline{g_3(F)}^{CDA'B'}\Big)\epsilon_{AB}\Bigg)\xi^B = 0 \, ,
\label{eq:general_YM_int_constraint_developed_valued}
\end{align}
which is equivalent to the desired vanishing of the auxiliary field $\xi_A=0$, except for some extreme values of $F$ when the matrix vanishes, but the perturbation theory becomes inadequate long before that. Let us remark that the first line of this expression looks like the complex conjugate of the left-hand side of  \eqref{eq:eq_g_2_g_3_(0,2)}, but with a twisted order of the functions in each term. It means that the first line vanishes for abelian interactions.

\section{Constant electromagnetic field}
Let us focus on the constant electromagnetic background. That it is electromagnetic means that we gauged $so(2)$ and, hence, $F_{AB}\equiv F_{AB}{}\fud{i}{j}\equiv F_{AB}{}\epsilon\fud{i}{j}$, idem for $F_{A'B'}$. Therefore, whenever two Lorentz indices are contracted $F_{AC}F\fdu{B}{C}=\tfrac12 \epsilon_{AB}F_{MN} F^{MN}$ we get a scalar $F^2=F_{MN} F^{MN}$, which is not true for a generic Yang-Mills interaction. ``Constant'' means $D_\mu F_{AB}=\pl_\mu F_{AB}=0$, idem for $F_{A'B'}$. Given that we now have only $\epsilon_{AB}$ and $F_{AB}$, which have opposite symmetries, we can constrain the structure functions further
\besubeqs\label{defg}
\begin{align}
g_2(F)_{AB} &= -\frac{1}{m}\Big(1-f_1(F^2,\bar{F}^2)\Big)F_{AB} \, ,
\label{eq:general_g_2(2,0)_dvlp_abelian_valued} \\
g_2(F) &= 6mf_2(F^2,\bar{F}^2) \, ,
\label{eq:general_g_2(0,0)_dvlp_abelian_valued} \\
g_3(F)_{ABA'B'} &= -\frac{1}{2m^3}\Big(1 + ib + f_3(F^2,\bar{F}^2)\Big)F_{AB}F_{A'B'} \, ,
\label{eq:general_g_3(2,2)_dvlp_abelian_valued}
\end{align}
\esubeqs
where $f_1$, $f_2$ and $f_3$ are arbitrary functions of $F^2$ and $\bar{F}^2$. Note that with the normalization above these functions are dimensionless and the equations are more elegant. With this ansatz, the algebraic equations \eqref{eq:eq_g_2_g_3_(0,2)} and \eqref{eq:eq_g_2_g_3_(2,2)} become, respectively,
\besubeqs
\begin{align}
C_1 := \frac{F^2}{4m^4}\big(1+ib\big) - \bar{f}_1 - f_2 + \bar{f}_1f_2 - \frac{F^2}{4m^4}\big(1+ib\big)f_1 + \frac{F^2}{4m^4}f_3 - \frac{F^2}{4m^4}f_1f_3 &= 0 \, ,
\label{eq:eq_funct_f_(1)} \\
C_2 := 2\big(f_1 + \bar{f}_1\big) - \big((1+ib)f_2 + (1-ib)\bar{f}_2\big) + \big(f_3 + \bar{f}_3\big) - 2f_1\bar{f}_1 - \big(f_2f_3 + \bar{f}_2\bar{f}_3\big) &= 0 \, .
\label{eq:eq_funct_f_(2)}
\end{align}
\esubeqs
The problem is therefore reduced to these two scalar algebraic equations. The first term of the first equation implies that $f_1 = f_2 = f_3 = 0$ is not a solution, showing why the first and the second order nonminimal interactions are not sufficient. Note also that $C_2 \equiv \bar{C}_2$.

In the abelian case, the equations of motion \eqref{eq:general_int_EOM_1_valued}, \eqref{eq:general_int_EOM_2_valued}, \eqref{eq:general_int_EOM_3_valued} and \eqref{eq:general_int_EOM_4_valued} become
{\allowdisplaybreaks
\begin{align}
{E^{\psi}}_{ABA'} &:= m\psi_{ABA'} + \sqrt{2}D_{(A}^{\phantom{(A}B'}\bar{\psi}_{B)A'B'} + \sqrt{2}D_{(A|A'|}\xi_{B)} - \frac{1}{2m^3}\Big(1+ib+f_3\Big)F_{A'B'}F_{C(A}\psi_{B)}^{\phantom{B)}CB'} \notag \\
&\quad\; + \frac{1}{m}\Big(1-\bar{f}_1\Big) F_{A'B'}\psi_{AB}^{\phantom{AB}B'} - m\bar{f}_2\psi_{ABA'} - \frac{1}{2m^3}\Big(1+ib+f_3\Big)F_{AB}F_{A'B'}\bar{\xi}^{B'} = 0 \, ,
\label{eq:general_int_abelian_EOM_1_valued} \\[3mm]
{E^{\bar{\psi}}}_{AA'B'} &:= -m\bar{\psi}_{AA'B'} + \sqrt{2}D^{C}_{\phantom{C}(A'}\psi_{|AC|B')} + \sqrt{2}D_{A(A'}\bar{\xi}_{B')} + \frac{1}{2m^3}\Big(1-ib+\bar{f}_3\Big)F_{AB}F_{C'(A'}\bar{\psi}_{\phantom{BC'}B')}^{BC'} \notag \\
&\quad\; - \frac{1}{m}\Big(1-f_1\Big)F_{AB}\bar{\psi}_{\phantom{B}A'B'}^{B} + mf_2\bar{\psi}_{AA'B'} + \frac{1}{2m^3}\Big(1-ib+\bar{f}_3\Big)F_{AB}F_{A'B'}\xi^{B} = 0 \, ,
\label{eq:general_int_abelian_EOM_2_valued} \\[3mm]
{E^{\xi}}_{A} &:= 6m\xi_{A} - 3\sqrt{2}D_{AA'}\bar{\xi}^{A'} - \sqrt{2}D^{CC'}\psi_{ACC'} - \frac{2}{m}\Big(1-f_1\Big)F_{AB}\xi^{B} \notag \\
&\quad\; - 6mf_2\xi_{A} - \frac{1}{2m^3}\Big(1-ib+\bar{f}_3\Big)F_{AB}F_{A'B'}\bar{\psi}^{BA'B'} = 0 \, ,
\label{eq:general_int_abelian_EOM_3_valued} \\[3mm]
{E^{\bar{\xi}}}_{A'} &:= -6m\bar{\xi}_{A'} - 3\sqrt{2}D_{AA'}\xi^{A} - \sqrt{2}D^{CC'}\bar{\psi}_{CC'A'} + \frac{2}{m}\Big(1-\bar{f}_1\Big)F_{A'B'}\bar{\xi}^{B'} \notag \\
&\quad\; + 6m\bar{f}_2\bar{\xi}_{A'} + \frac{1}{2m^3}\Big(1+ib+f_3\Big)F_{AB}F_{A'B'}\psi^{ABB'} = 0 \, .
\label{eq:general_int_abelian_EOM_4_valued}
\end{align}}
The constraint \eqref{eq:general_YM_int_constraint_valued} reduces to
\begin{align}
D^{BB'}{E^{\psi}}_{ABB'} + \frac{\sqrt{2}}{2}m{E^{\xi}}_{A} + &\frac{1}{2}D_{A}^{\phantom{A}A'}{E^{\bar{\xi}}}_{A'} +\frac{\sqrt{2}}{4m^3}\Big(1+ib+f_3\Big)F_{AB}F_{A'B'}{E^{\bar{\psi}}}^{BA'B'} \notag \\
&- \frac{\sqrt{2}}{2m}\Big(1-\bar{f}_1\Big)F_{A'B'}{E^{\bar{\psi}}}_{A}^{\phantom{A}A'B'} - \frac{\sqrt{2}}{2}m\bar{f}_2{E^{\xi}}_{A} = 0\, .
\label{eq:general_YM_int_abelian_constraint_valued}
\end{align}
With the help of the definitions \eqref{defg} the constraint \eqref{eq:general_YM_int_constraint_developed_valued} simplifies to
\begin{align}
\Bigg(1-f_2+\bar{f}_2 - f_2\bar{f}_2 - \frac{F^2\bar{F}^2}{48m^8}\Big(1+b^2+f_3+\bar{f}_3+ib\bar{f}_3-ibf_3+f_3\bar{f}_3\Big)\Bigg)&\xi_A - \frac{1}{3}\ \bar{C}_1\ F_{AB}\xi^B= 0 \, ,
\end{align}
where $C_1$ was defined in \eqref{eq:eq_funct_f_(1)}. The last term vanishes, in fact, which allows us to simplify it further
\begin{align}
\Bigg(1-f_2+\bar{f}_2 - f_2\bar{f}_2 - \frac{F^2\bar{F}^2}{48m^8}\Big(1+b^2+f_3+\bar{f}_3+ib\bar{f}_3-ibf_3+f_3\bar{f}_3\Big)\Bigg)\xi_A = 0 \, .
\end{align}
The final form of the constraint above ensures that the auxiliary field vanishes, save for some extreme values of $F^2$.

\subsection{Solution at low orders}
It is instructive to see how the solution of the algebraic constraints \eqref{eq:eq_funct_f_(1)} and \eqref{eq:eq_funct_f_(2)} look like at low orders. Below we expand $f_1$, $f_2$ and $f_3$ to the leading order
\besubeqs
\begin{align}
f_1(F^2,\bar{F}^2) &= a_{10}\frac{F^2}{4m^4} + a_{01}\frac{\bar{F}^2}{4m^4} + \mathcal{O}(F^4) \, ,
\label{eq:1st_order_expansion_f_1} \\
f_2(F^2,\bar{F}^2) &= b_{10}\frac{F^2}{4m^4} + b_{01}\frac{\bar{F}^2}{4m^4} + \mathcal{O}(F^4) \, ,
\label{eq:1st_order_expansion_f_2} \\
f_3(F^2,\bar{F}^2) &= c_{10}\frac{F^2}{4m^4} + c_{01}\frac{\bar{F}^2}{4m^4} + \mathcal{O}(F^4) \, .
\label{eq:1st_order_expansion_f_3}
\end{align}
\esubeqs
The equations are therefore satisfied, up to the first order in $F^2$, if and only if
\besubeqs
\begin{align}
b_{10} &= 1 + ib - \bar{a}_{01} \, ,
\label{eq:coeff_1st_order_F^2_1} \\
b_{01} &= -\bar{a}_{10} \, ,
\label{eq:coeff_1st_order_F^2_2} \\
c_{10} &= \big(1+ib\big)^2 - 3\big(a_{10} + \bar{a}_{01}\big) + ib\big(a_{10} - \bar{a}_{01}\big) - \bar{c}_{01} \, .
\label{eq:coeff_1st_order_F^2_3}
\end{align}
\esubeqs
The freedom in the interactions is given by a real parameter $b$ and three of the six complex parameters defining the functions $f$'s. By using this first order (in $F^2$ and $\bar{F}^2$) expansion of functions $f$'s, let us write the equations of motion \eqref{eq:general_int_abelian_EOM_1_valued} and \eqref{eq:general_int_abelian_EOM_3_valued} up to the third order in $F$
\begin{align}
m\psi_{ABA'} + \sqrt{2}D_{(A}^{\phantom{(A}B'}\bar{\psi}_{B)A'B'} + \sqrt{2}D_{(A|A'|}\xi_{B)} + \frac{1}{m}F_{A'B'}\psi_{AB}^{\phantom{AB}B'} - \frac{1}{2m^3}\big(1+ib\big)F_{(A|C}F_{A'B'}\psi_{|B)}^{\phantom{|B)}CB'}\quad & \notag \\
- \frac{1}{2m^3}\big(1+ib\big)F_{AB}F_{A'B'}\bar{\xi}^{B'} - \frac{1}{4m^3}\Big(\big(1+ib) - \bar{a}_{01}\Big)F^2\psi_{ABA'} + \frac{1}{4m^3}\bar{a}_{10}\bar{F}^2\psi_{ABA'}\quad & \notag \\
- \frac{1}{4m^5}a_{10}F^2F_{A'B'}\psi_{AB}^{\phantom{AB}B'} - \frac{1}{4m^5}a_{01}\bar{F}^2F_{A'B'}\psi_{AB}^{\phantom{AB}B'} = 0 \, , \qquad&
\label{eq:general_int_EOM_1_3rd_order} \\[3mm]
6m\xi_{A} - 3\sqrt{2}D_{AA'}\bar{\xi}^{A'} - \sqrt{2}D^{CC'}\psi_{ACC'} - \frac{2}{m}F_{AB}\xi^B - \frac{3}{2m^3}\Big(\big(1+ib)-\bar{a}_{01}\Big)F^2\xi_A + \frac{3}{2m^3}\bar{a}_{10}\bar{F}^2\xi_A  & \notag \\
- \frac{1}{2m^3}\big(1-ib\big)F_{AB}F_{A'B'}\bar{\psi}^{BA'B'} + \frac{1}{2m^5}a_{10}F^2F_{AB}\xi^B + \frac{1}{2m^5}a_{01}\bar{F}^2F_{AB}\xi^B = 0 \, .\qquad&
\label{eq:general_int_EOM_3_3rd_order}
\end{align}
As we knew already, the equations are completely fixed at the first order in $F$. The ambiguity pops up at the second order. It seems impossible to redefine the fields so that some free coefficients are absorbed. Therefore, starting from the second order we observe some nontrivial Wilson coefficients. Let us note that starting from the second order the constraint $\xi_A=0$ does not imply the transversality constraint for $\psi_{ABA'}$ but
\begin{align}
D^{CC'}\psi_{ACC'} = -\frac{\sqrt{2}}{4m^3}\big(1-ib+\bar{f}_3\big)F_{AB}F_{A'B'}\bar{\psi}^{BA'B'}\, ,
\label{eq:not_transverse_constraint}
\end{align}
which is obtained by setting $\xi_A=0$ in the equation of motion \eqref{eq:general_int_abelian_EOM_3_valued}. It is impossible to choose the coefficients to get the transversality constraint.

\subsection{Exact solution}
The main system of algebraic equations \eqref{eq:eq_funct_f_(1)}, \eqref{eq:eq_funct_f_(2)} admits plenty of solutions. In general, we expect infinitely many free Wilson coefficients that parameterize nonminimal interactions. There does not seem to exist polynomial solutions. Here, we will construct an exact solution.\footnote{This can be thought of as a further development of \cite{Porrati:2009bs}, where a certain system of the algebraic constraints was formulated to ensure that the auxiliary field decouples.} In view of the fact that selfdual fields, i.e. the ones where $F_{AB}=0$ or $F_{A'B'}=0$, play an important role in physics, let us assume that $f$'s depend either on $F^2$ or $\bar{F}^2$. Since function $f_1$ appears as $f_1$ and $\bar{f}_1$ in  \eqref{eq:eq_funct_f_(1)}, it may be easier to find such a solution if we assume $f_1=0$. In this case, Eq. \eqref{eq:eq_funct_f_(1)} becomes
\begin{align}
\frac{F^2}{4m^4}(1+ib) - f_2 + \frac{F^2}{4m^4}f_3 = 0 \, ,
\end{align}
which can be rewritten as
\begin{align}
f_3 = -\frac{1}{z}\Big((1+ib)z - f_2\Big) \, ,
\label{eq:exact_sol_f3_f2}
\end{align}
where we defined $z:= \frac{F^2}{4m^4}$ for a more compact notation. By plugging this expression into Eq. \eqref{eq:eq_funct_f_(2)} with $f_1 = 0$, we obtain
\begin{align}
(1+ib) - \frac{1}{z}f_2 + \frac{1}{z}f_2^2 + (1-ib) - \frac{1}{\bar{z}}\bar{f}_2 + \frac{1}{\bar{z}}\bar{f}_2^2 = 0 \, .
\end{align}
If we want that the functions are holomorphic, i.e. depend either on $z$ or $\bar{z}$, the first and the second halves of this equation can be solved independently, which leads to
\begin{align}
(1+ib) - \frac{1}{z}f_2 + \frac{1}{z}f_2^2 = ia \qquad \Leftrightarrow \qquad (1+i\Tilde{b})z - f_2 + f_2^2 = 0 \, ,
\end{align}
where $a$ and, hence, $\Tilde{b} := b - a$ are arbitrary real numbers. 
The solutions are
\begin{align}
\Bigg\{f_2 = \frac{1}{2}\Big(1-\sqrt{1-4(1+i\Tilde{b})z}\Big)\,,\quad \ f_2 = \frac{1}{2}\Big(1+\sqrt{1-4(1+i\Tilde{b})z}\Big)\Bigg\} \, .
\end{align}
However, by definition of the functions $f$'s, we need to satisfy $f_k(0,0) = 0$, for $k \in \{1,2,3\}$. Therefore, only the solution with the minus is physically acceptable here, i.e. we have finally
\besubeqs
\begin{align}
f_1\big(F^2, \bar{F}^2\big) &= 0 \, , \\
f_2\big(F^2, \bar{F}^2\big) &= \frac{1}{2}\Bigg(1 - \sqrt{1 - (1+i\Tilde{b})\frac{F^2}{m^4}}\Bigg) \, , \\
f_3\big(F^2, \bar{F}^2\big) &= \frac{4m^4}{F^2}\Bigg(1-(1+i\Tilde{b})\frac{F^2}{4m^4} - \sqrt{1 - (1+i\Tilde{b})\frac{F^2}{m^4}}\Bigg) \, .
\end{align}
\esubeqs
Consequently, we have found an exact solution to the algebraic equations that ensure vanishing of the auxiliary field.

\section{Chiralization}
As it was already mentioned in the introduction, a very efficient approach to constructing consistent interactions is the chiral approach. However, there is no efficient way to impose parity yet. Therefore, it is interesting to explore the relation between the chiral approach to massive higher-spin fields and the standard one where the physical field is in the $(s,s)$-representation of $sl(2,\mathbb{C})$ for bosons and in $(s-1/2,s+1/2)\oplus (s+1/2,s-1/2)$ for fermions.

\subsection{Chiralization in the free case}
The chiralization of the spin three-half at the free level and on Einstein backgrounds was discussed in \cite{Delplanque:2024enh}. Therefore, let us briefly recall the free case. Let us begin by considering the equations of motion \eqref{eq:spin3/2_free_EOM_2} and \eqref{eq:spin3/2_free_EOM_3}, respectively, as a definition of $\bar{\psi}_{AA'B'}$ and $\xi_A$. Then, we use these definitions in the two other equations of motion (\eqref{eq:spin3/2_free_EOM_1} and \eqref{eq:spin3/2_free_EOM_4}) in order to obtain the following second order equations of motion describing $\psi_{ABA'}$ and $\bar{\xi}_{A'}$
\besubeqs
\begin{align}
m\psi_{ABA'} - m^{-1}\Box\psi_{ABA'} + \tfrac{4}{3}m^{-1}\partial_{(A|A'|}\partial^{CC'}\psi_{B)CC'} &= 0 \, ,
\label{eq:spin3/2_free_EOM_(2,1)_2nd_order} \\
\bar{\xi}_{A'} &= 0 \, .
\end{align}
\esubeqs
The second equation is the ``suicide'' of the auxiliary field. The first one is a second order equation describing the main field $\psi_{ABA'}$. In order to obtain the chiral description, we define a new main field
\begin{align}
\varphi_{ABC} := m^{-1}\partial_{(A}^{\phantom{(A}A'}\psi_{BC)A'} \, .
\label{eq:spin3/2_free_def_(3,0)}
\end{align}
The definition allows us to rewrite the second order equation of motion as the first order one
\begin{align}
m\psi_{ABA'} + 2\partial^{C}_{\phantom{C}A'}\varphi_{ABC} = 0 \, .
\label{eq:spin3/2_free_EOM_(2,1)_(3,0)}
\end{align}
Finally, in order to obtain the chiral description, we swap the roles of the first order equations \eqref{eq:spin3/2_free_EOM_(2,1)_(3,0)} and the definition \eqref{eq:spin3/2_free_def_(3,0)}: Eq. \eqref{eq:spin3/2_free_EOM_(2,1)_(3,0)} becomes the definition of $\psi_{ABA'}$ in terms of $\varphi_{ABC}$ and the definition of $\varphi_{ABC}$ becomes the first order equation of motion. In doing so we obtain the following second order equations of motion
\begin{align}
\big(\Box - m^2\big)\varphi_{ABC} = 0 \, ,
\end{align}
which is the desired Klein-Gordon equation describing a massive spin-$3/2$ field in the chiral approach. The corresponding Lagrangian density is simply
\begin{align}
\mathcal{L} = \tfrac{1}{2}\varphi^{ABC}\big(\Box-m^2\big)\varphi_{ABC} \, .
\end{align}
In \cite{Delplanque:2024enh} we also checked that the transversality constraint for the old field is automatically satisfied once it is expressed in terms of the chiral one. 

\subsection{Chiralization in a constant electromagnetic field}
The Lagrangian density of the massive spin three-half field contains a lot of nonminimal terms. Therefore, let us apply the procedure of chiralization only at the first order in $F$. The Lagrangian density in this case is \eqref{eq:Lagrangian_1st_order_int_with_valued_coeff}, which leads to the following equations of motion
{\allowdisplaybreaks
\besubeqs
\begin{align}
{E^{\psi}}_{ABA'} &:= m\psi_{ABA'} + \sqrt{2}D_{(A}^{\phantom{(A}B'}\bar{\psi}_{B)A'B'} + \sqrt{2}D_{(A|A'|}\xi_{B)} + \frac{1}{m}F_{A'B'}\psi_{AB}^{\phantom{AB}B'} + \mathcal{O}(F^2) = 0 \, ,
\label{eq:spin3/2_EM_EOM_1_1st_order} \\[3mm]
{E^{\bar{\psi}}}_{AA'B'} &:= -m\bar{\psi}_{AA'B'} + \sqrt{2}D^{C}_{\phantom{C}(A'}\psi_{|AC|B')} + \sqrt{2}D_{A(A'}\bar{\xi}_{B')} - \frac{1}{m}F_{AB}\bar{\psi}_{\phantom{B}A'B'}^{B} + \mathcal{O}(F^2) = 0 \, ,
\label{eq:spin3/2_EM_EOM_2_1st_order} \\[3mm]
{E^{\xi}}_{A} &:= 6m\xi_{A} - 3\sqrt{2}D_{AA'}\bar{\xi}^{A'} - \sqrt{2}D^{CC'}\psi_{ACC'} - \frac{2}{m}F_{AB}\xi^B + \mathcal{O}(F^2) = 0 \, ,
\label{eq:spin3/2_EM_EOM_3_1st_order} \\[3mm]
{E^{\bar{\xi}}}_{A'} &:= -6m\bar{\xi}_{A'} - 3\sqrt{2}D_{AA'}\xi^{A} - \sqrt{2}D^{CC'}\bar{\psi}_{CC'A'} + \frac{2}{m}F_{A'B'}\bar{\xi}^{B'} + \mathcal{O}(F^2) = 0 \, .
\label{eq:spin3/2_EM_EOM_4_1st_order}
\end{align}
\esubeqs}%
The equations \eqref{eq:spin3/2_EM_EOM_2_1st_order} and \eqref{eq:spin3/2_EM_EOM_3_1st_order} can respectively be rewritten as
\besubeqs
\begin{align}
\bar{\psi}_{AA'B'} &= \frac{\sqrt{2}}{m}\Big(D^{C}_{\phantom{C}(A'|}\psi_{AC|B')} + D_{A(A'}\bar{\xi}_{B')}\Big) + \frac{\sqrt{2}}{m^3}\Big(F_{A}^{\phantom{A}B}D^{C}_{\phantom{C}(A'|}\psi_{BC|B')} + F_{A}^{\phantom{A}B}D_{B(A'}\bar{\xi}_{B')}\Big) + \mathcal{O}(F^2) \, ,
\label{eq:spin3/2_EM_EOM_2_1st_order_psi_bar_isolated} \\
\xi_A &= \frac{\sqrt{2}}{2m}\Big(D_{AA'}\bar{\xi}^{A'} + \tfrac{1}{3}D^{CC'}\psi_{ACC'}\Big) - \frac{\sqrt{2}}{6m^3}\Big(F_{A}^{\phantom{A}B}D_{BA'}\bar{\xi}^{A'} + \tfrac{1}{3}F_{A}^{\phantom{A}B}D^{CC'}\psi_{BCC'}\Big) + \mathcal{O}(F^2) \, .
\label{eq:spin3/2_EM_EOM_3_1st_order_xi_isolated}
\end{align}
\esubeqs
By using these equations \eqref{eq:spin3/2_EM_EOM_2_1st_order_psi_bar_isolated} and \eqref{eq:spin3/2_EM_EOM_3_1st_order_xi_isolated} as definitions of $\bar{\psi}_{AA'B'}$ and $\xi_A$ in terms of $\psi_{ABA'}$ and $\bar{\xi}$ inside Eqs. \eqref{eq:spin3/2_EM_EOM_1_1st_order} and \eqref{eq:spin3/2_EM_EOM_4_1st_order}, we obtain the second order equations for $\psi_{ABA'}$ and $\bar{\xi}$
\besubeqs
\begin{align}
&m^2\psi_{ABA'} - \Box\psi_{ABA'} + \frac{4}{3}D_{(A|A'}D^{CC'}\psi_{|B)CC'} \notag \\
&\, - F_{(A}^{\phantom{(A}D}\psi_{B)DA'} + F_{A'B'}\psi_{AB}^{\phantom{AB}B'} - F_{AB}\bar{\xi}_{A'} -\frac{1}{m^2}F_{(A}^{\phantom{(A}C}\Box\psi_{B)CA'} \notag \\
&\, + \frac{8}{9m^2}F_{(A}^{\phantom{(A}D}D_{B)A'}D^{CC'}\psi_{DCC'} - \frac{4}{3m^2}F_{(A}^{\phantom{(A}C}D_{B)A'}D_{CB'}\bar{\xi}^{B'} + \frac{1}{m^2}F_{AB}\Box\bar{\xi}_{A'} + \mathcal{O}(F^2) = 0 \, ,
\label{eq:spin3/2_EM_EOM_(2,1)_2nd_order_before} \\[5mm]
&\bar{\xi}_{A'} -\frac{1}{6m}\Big(m^2\psi_{ABA'} - \Box\psi_{ABA'} + \frac{4}{3}D_{AA'}D^{CC'}\psi_{BCC'}\Big) + \mathcal{O}(F^2) = 0 \, .
\end{align}
\esubeqs
With the help of the first equation of motion, we can rewrite the second one as
\begin{align}
\bar{\xi}_{A'} + \mathcal{O}(F^2) = 0 \, ,
\label{eq:spin3/2_EM_EOM_(0,1)_2nd_order}
\end{align}
which is the suicide of the auxiliary field to the required order. By using this last equation in the first one \eqref{eq:spin3/2_EM_EOM_(2,1)_2nd_order_before}, it becomes
\begin{align}
m^2\psi_{ABA'} - \Box\psi_{ABA'} &+ \frac{4}{3}D_{(A|A'}D^{CC'}\psi_{|B)CC'} - F_{(A}^{\phantom{(A}D}\psi_{B)DA'} + F_{A'B'}\psi_{AB}^{\phantom{AB}B'} \notag \\
&- \frac{1}{m^2}F_{(A}^{\phantom{(A}C}\Box\psi_{B)CA'} + \frac{8}{9m^2}F_{(A}^{\phantom{(A}D}D_{B)A'}D^{CC'}\psi_{DCC'} + \mathcal{O}(F^2) = 0 \, .
\label{eq:spin3/2_EM_EOM_(2,1)_2nd_order}
\end{align}
Let us continue the chiralization procedure by defining the following new chiral field
\begin{align}
\varphi_{ABC} := m^{-1}D_{(A}^{\phantom{(A}C'}\psi_{BC)C'} + \frac{1}{3}(2a+1)m^{-3}F_{(AB}D^{DD'}\psi_{C)DD'} - am^{-3}F_{D(A}D^{DC'}\psi_{BC)C'} + \mathcal{O}(F^2) \, ,
\label{eq:EM_def_(3,0)}
\end{align}
where $a$ is an arbitrary complex number. We can then rewrite the equations of motion for $\psi_{ABA'}$ \eqref{eq:spin3/2_EM_EOM_(2,1)_2nd_order} as 
\begin{align}
\psi_{ABA'} = -2m^{-1}D^{C}_{\phantom{C}A'}\varphi_{ABC} - \frac{2}{3}am^{-3}F^{CD}D_{DA'}\varphi_{ABC} + \frac{4}{3}(a-3)m^{-3}F_{(A|}^{\phantom{(A|}C}D^{D}_{\phantom{D}A'}\varphi_{|B)CD} + \mathcal{O}(F^2) \, .
\label{eq:EM_(2,0)_(3,0)}
\end{align}
The final step of chiralization consists in replacing  $\psi_{ABA'}$ in \eqref{eq:EM_def_(3,0)} with the help of the relation \eqref{eq:EM_(2,0)_(3,0)}. It gives the equations of motion expressed in terms of the chiral field
\begin{align}
\big(\Box - m^2\big)\varphi_{ABC} + 3F_{(A}^{\phantom{(A}D}\varphi_{BC)D} + \mathcal{O}(F^2) = 0 \, .
\label{eq:EM_EOM_(3,0)}
\end{align}
Let us note that even though the parameter $a$ is completely arbitrary, it does not have any physical effect because it disappears in the equations of motion. In fact, some of the steps of the chiralization can be simplified since the physical field $\psi_{ABA'}$ is transverse to the required order, which eliminates the last term in \eqref{eq:spin3/2_EM_EOM_(2,1)_2nd_order}. Also, we can use the free equations of motion for the $F\square \psi$-term in \eqref{eq:spin3/2_EM_EOM_(2,1)_2nd_order}. It is then obvious that the wave equation contains the standard D'Alembert operator and there is no acausal propagation. The final equations can be obtained from the following Lagrangian density
\begin{align}
\mathcal{L} = \frac{1}{2}\varphi^{ABC}\big(\Box - m^2\big)\varphi_{ABC} + \frac{3}{2}\varphi^{ABC}F_{A}^{\phantom{A}D}\varphi_{BCD} + \mathcal{O}(F^2) \, .
\label{eq:EM_Lagrangian_(3,0)}
\end{align}
The Lagrangian coincides with the one in \cite{Cangemi:2023ysz} for the so-called root-Kerr theory in the chiral approach, provided that we remember that $\pl_\mu F=0$, and we can use the free equations to simplify the structure of the cubic terms which collapse into a single one above. One can still talk about the gyromagnetic ratio $g$ in the chiral approach (in general, it can be split into left and right and we, obviously, have only one of them). It is clear that $g=2s$. 

\subsubsection{The fate of the constraints}
Let us check how the constraints transform during the chiralization procedure. Since the only sensible equation that the chiral field can satisfy is the Klein-Gordon equation with, possibly, nonminimal terms, other constraints, e.g. the transversality, must be satisfied automatically when the old field variables are expressed in terms of the chiral field. The first step of the chiralization consists in passing from the set of Eqs. \eqref{eq:spin3/2_EM_EOM_1_1st_order}, \eqref{eq:spin3/2_EM_EOM_2_1st_order}, \eqref{eq:spin3/2_EM_EOM_3_1st_order} and \eqref{eq:spin3/2_EM_EOM_4_1st_order} describing the four fields $\psi_{ABA'}$, $\bar{\psi}_{AA'B'}$, $\xi_A$ and $\bar{\xi}_{A'}$, to the set \eqref{eq:spin3/2_EM_EOM_(0,1)_2nd_order}, \eqref{eq:spin3/2_EM_EOM_(2,1)_2nd_order} describing the two fields $\psi_{ABA'}$ and $\bar{\xi}_{A'}$. The constraints we can extract from the first set are
\besubeqs
\begin{align}
D^{CC'}\psi_{ACC'} &= 0 \, , \\
D^{CC'}\bar{\psi}_{CC'A'} &= 0 \, , \\
\xi_A &= 0 \, , \\
\bar{\xi}_{A'} &= 0 \, .
\end{align}
\esubeqs
The first and fourth ones are also the constraints we can extract from the new set of equations of motion. Let us check that the second and third constraints become trivial when expressed in terms of the fields of the new set of equations of motion. Let us begin with the third constraint. By using the relation \eqref{eq:spin3/2_EM_EOM_3_1st_order_xi_isolated}, we can rewrite it as
\begin{align}
\frac{\sqrt{2}}{2m}\Big(D_{AA'}\bar{\xi}^{A'} + \tfrac{1}{3}D^{CC'}\psi_{ACC'}\Big) - \frac{\sqrt{2}}{6m^3}\Big(F_{A}^{\phantom{A}B}D_{BA'}\bar{\xi}^{A'} + \tfrac{1}{3}F_{A}^{\phantom{A}B}D^{CC'}\psi_{BCC'}\Big) + \mathcal{O}(F^2) = 0 \, .
\end{align}
This relation is trivially satisfied because of the constraints that follow from the new set of equations of motion (\emph{i.e.} the first and fourth constraints here). Let us check the second constraint (the conjugate transversality constraint $D^{CC'}\bar{\psi}_{CC'A'} = 0$). By using the relation \eqref{eq:spin3/2_EM_EOM_2_1st_order_psi_bar_isolated}, we can rewrite this constraint as
\begin{align}
\frac{\sqrt{2}}{m}\Big(D^{AA'}D^{C}_{\phantom{C}(A'|}\psi_{AC|B')} &+ D^{AA'}D_{A(A'}\bar{\xi}_{B')}\Big) \notag \\
&+ \frac{\sqrt{2}}{m^3}\Big(F_{A}^{\phantom{A}B}D^{AA'}D^{C}_{\phantom{C}(A'|}\psi_{BC|B')} + F_{A}^{\phantom{A}B}D^{AA'}D_{B(A'}\bar{\xi}_{B')}\Big) + \mathcal{O}(F^2) = 0 \, , \notag \\
{ } \notag \\
\Leftrightarrow &\quad F^{AB}\big(\Box - m^2\big)\psi_{ABB'} + \mathcal{O}(F^2) = 0 \, ,
\end{align}
where the last line is obtained by applying the constraints on $\psi_{ABA'}$ and $\bar{\xi}_{A'}$. This expression is trivially satisfied according to the equations of motion for $\psi_{ABA'}$ \eqref{eq:spin3/2_EM_EOM_(2,1)_2nd_order}. Now, let us check that the constraints become trivial at the second (and last) step of the chiralization. In this last step we pass from the set of Eqs. \eqref{eq:spin3/2_EM_EOM_(0,1)_2nd_order}, and \eqref{eq:spin3/2_EM_EOM_(2,1)_2nd_order} describing the fields $\psi_{ABA'}$ and $\bar{\xi}_{A'}$ , to the equation of motion \eqref{eq:EM_EOM_(3,0)} describing the field $\varphi_{ABC}$. The vanishing of the auxiliary field $\bar{\xi}_{A'}$ appears directly as an equation of motion and we do not have to consider it anymore. The last constraint we need to check is the transversality constraint on the field $\psi_{ABA'}$. By using the expression \eqref{eq:EM_(2,0)_(3,0)}, we can rewrite the transversality constraint in terms of the new field. We obtain
\begin{align}
-2m^{-1}D^{BA'}D^{C}_{\phantom{C}A'}\varphi_{ABC} -& \frac{2}{3}am^{-3}F^{CD}D^{BA'}D_{DA'}\varphi_{ABC} + \notag\\
&+\frac{4}{3}(a-3)m^{-3}F_{(A|}^{\phantom{(A|}C}D^{BA'}D^{D}_{\phantom{D}A'}\varphi_{|B)CD} + \mathcal{O}(F^2) = 0 \, , \notag \\
\Leftrightarrow \quad &F^{BC}\big(\Box - m^2\big)\varphi_{ABC} + \mathcal{O}(F^2) = 0 \, ,
\end{align}
which is trivially satisfied according to the chiral equations of motion \eqref{eq:EM_EOM_(3,0)}.

\section{Conclusions and Discussion}
\label{sec:}
In this paper, we have found the system of two algebraic constraints that are equivalent to the vanishing of the auxiliary fields $\xi_A$, $\bar\xi_{A'}$ in the Rarita-Schwinger action coupled to a constant electromagnetic/Yang-Mills background. For the case of a constant electromagnetic background we have also found a simple exact solution to the system, which is nonpolynomial. It is also clear that the transversality constraint gets modified starting from the second order in $F$. However, it is not clear if the latter is a sign of any problem. For example, in \cite{Chiodaroli:2021eug,Cangemi:2022bew,Cangemi:2023ysz} a theory that couples a massive spin-$s$ field to electromagnetic/Yang-Mills field was constructed up to the quartic order and for dynamical (nonconstant) electromagnetic/YM fields and it does not reveal any pathology. 

We have not really explored the genuine Yang-Mills interactions in this paper, which would be interesting to do in the future, in particular, to look for exact solutions. Another deformation direction to turn on is to allow for non-constant backgrounds. The simplest type of non-constant backgrounds are self-dual configurations aka instantons. It is likely that the solution will depend on all derivatives of a self-dual Yang-Mills field, the relations among which can nicely be encoded by a strong homotopy algebra found in \cite{Skvortsov:2022unu}. It would also be important to perform the chiralization at all orders, which should teach us how parity in the standard approach transmutates into a specific set of nonminimal interactions in the chiral approach.

\section*{Acknowledgments}
\label{sec:Aknowledgements}
The work of W.D. and E. S. was partially supported by the European Research Council (ERC) under the European Union’s Horizon 2020 research and innovation programme (grant agreement No 101002551). The work of W.D. was also supported by UMONS stipend ``Bourse d'encouragement doctorale FRIA/FRESH''. E.S. is grateful to Nicolas Boulanger, Maxim Grigoriev, Alexander Ochirov, Alexey Sharapov, Mirian Tsulaia and Yuri Zinoviev for many useful discussions on the topic.

\appendix

\section{The constraint}
\label{app:bigmess}
The constraint \eqref{eq:general_YM_int_constraint_decomposed} acquires the following final form (we need to use the Fierz identities sometimes)
{\allowdisplaybreaks
\begin{align}
\Bigg(\Big(-& 3\sqrt{2}m^2 + \frac{\sqrt{2}}{2}mg_2(F) - 3mh_2(F) + \frac{1}{2}h_2(F)g_2(F) - \frac{1}{4}h_1(F)_{A'B'}\overline{g_3(F)}^{A'B'} + h_2(F)_{CD}g_2(F)^{CD} \notag \\
-& \frac{1}{2}h_1(F)_{CDA'B'}\overline{g_3(F)}^{CDA'B'}\Big)\epsilon_{AB} + \Big(\sqrt{2}F_{AB} + \sqrt{2}mg_2(F)_{AB} + 6mh_2(F)_{AB} - h_2(F)_{AB}g_2(F) \notag \\
-& h_2(F)g_2(F)_{AB} + \frac{1}{2}h_1(F)_{ABA'B'}\overline{g_3(F)}^{A'B'} + \frac{1}{2}h_1(F)_{A'B'}\overline{g_3(F)}_{AB}^{\phantom{AB}A'B'} + 2h_2(F)_{C(A}g_2(F)_{B)}^{\phantom{B)}C} \notag \\
-& h_1(F)_{(A|CA'B'}\overline{g_3(F)}_{|B)}^{\phantom{|B)}CA'B'}\Big)\Bigg)\xi^B \notag \\
&\hspace{-30pt} + \Big(\sqrt{2}h_1(F)_{ABA'B'} + g_3(F)_{ABA'B'}\Big)D^{BB'}\bar{\xi}^{A'} - \Big(\overline{g_2(F)}_{A'B'} + \frac{\sqrt{2}}{2}h_1(F)_{A'B'}\Big)D_A^{\phantom{A}A'}\bar{\xi}^{B'} \notag \\
&\hspace{-30pt} + \Big(3\sqrt{2}h_2(F)_{AB} - \frac{1}{2}g_3(F)_{AB}\Big)D^{BB'}\bar{\xi}_{B'} + \Big(\frac{3\sqrt{2}}{2}h_2(F) - \frac{1}{2}\overline{g_2(F)}\Big)D_{AA'}\bar{\xi}^{A'} \notag \\
&\hspace{-30pt} + 2g_1(F)_{ABCDB'C'}D^{BB'}\psi^{CDC'} - g_1(F)_{ABCD}D^{BB'}\psi^{CD}_{\phantom{CD}B'} \notag \\
&\hspace{-30pt} + \Big(\frac{\sqrt{2}}{2}h_2(F) - \frac{1}{3}g_1(F)\Big)D^{BB'}\psi_{ABB'} + \Big(g_1(F)_{AB} - \sqrt{2}h_2(F)_{AB}\Big)D^{CC'}\psi^{B}_{\phantom{B}CC'} \notag \\
&\hspace{-30pt} - \Big(g_1(F)_{BCA'B'} + \frac{1}{2}g^T_3(F)_{BCA'B'}\Big)D_A^{\phantom{A}A'}\psi^{BCB'} - \frac{1}{2}\Big(g_1(F)_{BC} - \frac{1}{2}g^T_3(F)_{BC}\Big)D_{AA'}\psi^{BCA'} \notag \\
&\hspace{-30pt} + \Big(\sqrt{2}h_1(F)_{ACA'B'} - 2g_1(F)_{ACA'B'}\Big)D^{BA'}\psi_{B}^{\phantom{B}CB'} + \Big(\frac{2}{3}g_1(F)_{A'B'} - \frac{\sqrt{2}}{2}h_1(F)_{A'B'}\Big)D^{BA'}\psi_{AB}^{\phantom{AB}B'} \notag \\
&\hspace{-30pt} - \Bigg(\frac{\sqrt{2}}{2}F_{A'B'} - \frac{\sqrt{2}}{4}m\overline{g^T_3(F)}_{A'B'} - \frac{1}{2}mh_1(F)_{A'B'} + \frac{1}{2}h_1(F)_{C'D'}\overline{g_1(F)}^{C'D'}_{\phantom{C'D'}A'B'} - \frac{1}{2}h_1(F)_{B'C'}\overline{g_1(F)}_{A'}^{\phantom{A'}C'} \notag \\
+& \frac{1}{6}h_1(F)_{A'B'}\overline{g_1(F)} + \frac{1}{4}h_2(F)\overline{g^T_3(F)}_{A'B'} + \frac{1}{2}h_2(F)_{AC}\overline{g^T_3(F)}^{C}_{\phantom{C}BA'B'} - \frac{1}{3}h_1(F)_{CDA'B'}\overline{g_1(F)}^{CD} \notag \\
+& h_1(F)_{CDB'C'}\overline{g_1(F)}^{CDC'}_{\phantom{CDC'}A'} - h_1(F)^{BCC'D'}\overline{g_1(F)}_{BCA'B'C'D'}\Bigg)\bar{\psi}_A^{\phantom{A}A'B'} \notag \\
&\hspace{-30pt} + \Bigg(\frac{\sqrt{2}}{2}m\overline{g^T_3(F)}_{ABA'B'} - mh_1(F)_{ABA'B'} - \frac{1}{2}h_2(F)\overline{g^T_3(F)}_{ABA'B'} + \frac{1}{2}h_2(F)_{AB}\overline{g^T_3(F)}_{A'B'} \notag \\
+& \frac{1}{3}h_1(F)_{A'B'}\overline{g_1(F)}_{AB} + \frac{1}{3}h_1(F)_{ABA'B'}\overline{g_1(F)} - h_1(F)_{ABB'C'}\overline{g_1(F)}_{A'}^{\phantom{A'}C'} + h_1(F)^{C'D'}\overline{g_1(F)}_{ABA'B'C'D'} \notag \\
-& h_1(F)_{B'C'}\overline{g_1(F)}_{ABA'}^{\phantom{ABA'}C'} + h_1(F)_{ABC'D'}\overline{g_1(F)}_{A'B'}^{\phantom{A'B'}C'D'} + h_2(F)_{C(A}\overline{g^T_3(F)}_{B)\phantom{C}A'B'}^{\phantom{B)}C} \notag \\
-& \frac{2}{3}h_1(F)_{(A|CA'B'}\overline{g_1(F)}_{|B)}^{\phantom{|B)}C} + h_1(F)_{(A|CB'C'}\overline{g_1(F)}_{|B)\phantom{C}A'}^{\phantom{|B)}C\phantom{A'}C'} + \notag\\
&+2h_1(F)_{(A}^{\phantom{(A}CC'D'}\overline{g_1(F)}_{B)CA'B'C'D'}\Bigg)\bar{\psi}^{BA'B'} = 0 \, .
\label{eq:general_YM_int_constraint_developed}
\end{align}}

\footnotesize
\providecommand{\href}[2]{#2}\begingroup\raggedright\endgroup


\begin{thebibliography}{10}

\bibitem{Wigner:1939cj}
E.~P. Wigner, ``{On Unitary Representations of the Inhomogeneous Lorentz Group},'' \href{http://dx.doi.org/10.2307/1968551}{{\em Annals Math.} {\bfseries 40} (1939) 149--204}.
[Reprint: Nucl. Phys. Proc. Suppl.6,9(1989)].

\bibitem{Bekaert:2006py}
X.~Bekaert and N.~Boulanger, ``The unitary representations of the poincare group in any spacetime dimension,''
\href{http://arxiv.org/abs/hep-th/0611263}{{\ttfamily hep-th/0611263}}.

\bibitem{Bekaert:2017khg}
X.~Bekaert and E.~D. Skvortsov, ``{Elementary particles with continuous spin},'' \href{http://dx.doi.org/10.1142/S0217751X17300198}{{\em Int. J. Mod. Phys. A} {\bfseries 32} no.~23n24, (2017) 1730019}, \href{http://arxiv.org/abs/1708.01030}{{\ttfamily arXiv:1708.01030 [hep-th]}}.

\bibitem{Bergshoeff:2009hq}
E.~A. Bergshoeff, O.~Hohm, and P.~K. Townsend, ``{Massive Gravity in Three Dimensions},'' \href{http://dx.doi.org/10.1103/PhysRevLett.102.201301}{{\em Phys. Rev. Lett.} {\bfseries 102} (2009) 201301}, \href{http://arxiv.org/abs/0901.1766}{{\ttfamily arXiv:0901.1766 [hep-th]}}.

\bibitem{deRham:2010kj}
C.~de~Rham, G.~Gabadadze, and A.~J. Tolley, ``{Resummation of Massive Gravity},'' \href{http://dx.doi.org/10.1103/PhysRevLett.106.231101}{{\em Phys. Rev. Lett.} {\bfseries 106} (2011) 231101}, \href{http://arxiv.org/abs/1011.1232}{{\ttfamily arXiv:1011.1232 [hep-th]}}.

\bibitem{Hassan:2011zd}
S.~F. Hassan and R.~A. Rosen, ``{Bimetric Gravity from Ghost-free Massive Gravity},'' \href{http://dx.doi.org/10.1007/JHEP02(2012)126}{{\em JHEP} {\bfseries 02} (2012) 126}, \href{http://arxiv.org/abs/1109.3515}{{\ttfamily arXiv:1109.3515 [hep-th]}}.

\bibitem{deRham:2014zqa}
C.~de~Rham, ``{Massive Gravity},'' \href{http://dx.doi.org/10.12942/lrr-2014-7}{{\em Living Rev. Rel.} {\bfseries 17} (2014) 7}, \href{http://arxiv.org/abs/1401.4173}{{\ttfamily arXiv:1401.4173 [hep-th]}}.

\bibitem{Bekaert:2022poo}
X.~Bekaert, N.~Boulanger, A.~Campoleoni, M.~Chiodaroli, D.~Francia, M.~Grigoriev, E.~Sezgin, and E.~Skvortsov, ``{Snowmass White Paper: Higher Spin Gravity and Higher Spin Symmetry},'' \href{http://arxiv.org/abs/2205.01567}{{\ttfamily arXiv:2205.01567 [hep-th]}}.

\bibitem{Boulware:1972yco}
D.~G. Boulware and S.~Deser, ``{Can gravitation have a finite range?},'' \href{http://dx.doi.org/10.1103/PhysRevD.6.3368}{{\em Phys. Rev. D} {\bfseries 6} (1972) 3368--3382}.

\bibitem{Velo:1969bt}
G.~Velo and D.~Zwanziger, ``{Propagation and quantization of Rarita-Schwinger waves in an external electromagnetic potential},''
\href{http://dx.doi.org/10.1103/PhysRev.186.1337}{{\em Phys. Rev.} {\bfseries 186} (1969) 1337--1341}.

\bibitem{Fierz:1939ix}
M.~Fierz and W.~Pauli, ``{On relativistic wave equations for particles of arbitrary spin in an electromagnetic field},'' \href{http://dx.doi.org/10.1098/rspa.1939.0140}{{\em Proc. Roy. Soc. Lond. A} {\bfseries 173} (1939) 211--232}.

\bibitem{Singh:1974qz}
L.~P.~S. Singh and C.~R. Hagen, ``{Lagrangian formulation for arbitrary spin. 1. The boson case},''
\href{http://dx.doi.org/10.1103/PhysRevD.9.898}{{\em Phys. Rev.} {\bfseries D9} (1974) 898--909}.

\bibitem{Singh:1974rc}
L.~P.~S. Singh and C.~R. Hagen, ``{Lagrangian formulation for arbitrary spin. 2. The fermion case},''
\href{http://dx.doi.org/10.1103/PhysRevD.9.910}{{\em Phys. Rev.} {\bfseries D9} (1974) 910--920}.

\bibitem{Pashnev:1989gm}
A.~I. Pashnev, ``{Composite Systems and Field Theory for a Free Regge Trajectory},'' \href{http://dx.doi.org/10.1007/BF01017664}{{\em Theor. Math. Phys.} {\bfseries 78} (1989) 272--277}.

\bibitem{Zinoviev:2001dt}
Y.~M. Zinoviev, ``{On massive high spin particles in AdS},'' \href{http://arxiv.org/abs/hep-th/0108192}{{\ttfamily arXiv:hep-th/0108192}}.

\bibitem{Zinoviev:2006im}
Y.~M. Zinoviev, ``{On massive spin 2 interactions},'' \href{http://dx.doi.org/10.1016/j.nuclphysb.2007.02.005}{{\em Nucl. Phys. B} {\bfseries 770} (2007) 83--106}, \href{http://arxiv.org/abs/hep-th/0609170}{{\ttfamily arXiv:hep-th/0609170}}.

\bibitem{Zinoviev:2008ck}
Y.~M. Zinoviev, ``{On spin 3 interacting with gravity},'' \href{http://dx.doi.org/10.1088/0264-9381/26/3/035022}{{\em Class. Quant. Grav.} {\bfseries 26} (2009) 035022}, \href{http://arxiv.org/abs/0805.2226}{{\ttfamily arXiv:0805.2226 [hep-th]}}.

\bibitem{Zinoviev:2009hu}
Y.~M. Zinoviev, ``{On massive spin 2 electromagnetic interactions},'' \href{http://dx.doi.org/10.1016/j.nuclphysb.2009.04.027}{{\em Nucl. Phys. B} {\bfseries 821} (2009) 431--451}, \href{http://arxiv.org/abs/0901.3462}{{\ttfamily arXiv:0901.3462 [hep-th]}}.

\bibitem{Zinoviev:2010cr}
Y.~M. Zinoviev, ``{Spin 3 cubic vertices in a frame-like formalism},'' \href{http://dx.doi.org/10.1007/JHEP08(2010)084}{{\em JHEP} {\bfseries 08} (2010) 084}, \href{http://arxiv.org/abs/1007.0158}{{\ttfamily arXiv:1007.0158 [hep-th]}}.

\bibitem{Buchbinder:2012iz}
I.~L. Buchbinder, T.~V. Snegirev, and Y.~M. Zinoviev, ``{Cubic interaction vertex of higher-spin fields with external electromagnetic field},'' \href{http://dx.doi.org/10.1016/j.nuclphysb.2012.07.012}{{\em Nucl. Phys. B} {\bfseries 864} (2012) 694--721}, \href{http://arxiv.org/abs/1204.2341}{{\ttfamily arXiv:1204.2341 [hep-th]}}.

\bibitem{Ochirov:2022nqz}
A.~Ochirov and E.~Skvortsov, ``{Chiral Approach to Massive Higher Spins},'' \href{http://dx.doi.org/10.1103/PhysRevLett.129.241601}{{\em Phys. Rev. Lett.} {\bfseries 129} no.~24, (2022) 241601}, \href{http://arxiv.org/abs/2207.14597}{{\ttfamily arXiv:2207.14597 [hep-th]}}.

\bibitem{Delplanque:2024enh}
W.~Delplanque and E.~Skvortsov, ``{Symmetric vs. chiral approaches to massive fields with spin},'' \href{http://arxiv.org/abs/2405.13706}{{\ttfamily arXiv:2405.13706 [hep-th]}}.

\bibitem{Buchbinder:2005ua}
I.~L. Buchbinder and V.~A. Krykhtin, ``{Gauge invariant Lagrangian construction for massive bosonic higher spin fields in D dimensions},'' \href{http://dx.doi.org/10.1016/j.nuclphysb.2005.07.035}{{\em Nucl. Phys. B} {\bfseries 727} (2005) 537--563}, \href{http://arxiv.org/abs/hep-th/0505092}{{\ttfamily arXiv:hep-th/0505092}}.

\bibitem{Buchbinder:2007ix}
I.~L. Buchbinder, V.~A. Krykhtin, and H.~Takata, ``{Gauge invariant Lagrangian construction for massive bosonic mixed symmetry higher spin fields},'' \href{http://dx.doi.org/10.1016/j.physletb.2007.09.033}{{\em Phys. Lett. B} {\bfseries 656} (2007) 253--264}, \href{http://arxiv.org/abs/0707.2181}{{\ttfamily arXiv:0707.2181 [hep-th]}}.

\bibitem{Kaparulin:2012px}
D.~S. Kaparulin, S.~L. Lyakhovich, and A.~A. Sharapov, ``{Consistent interactions and involution},'' \href{http://dx.doi.org/10.1007/JHEP01(2013)097}{{\em JHEP} {\bfseries 01} (2013) 097}, \href{http://arxiv.org/abs/1210.6821}{{\ttfamily arXiv:1210.6821 [hep-th]}}.

\bibitem{Kazinski:2005eb}
P.~O. Kazinski, S.~L. Lyakhovich, and A.~A. Sharapov, ``{Lagrange structure and quantization},'' \href{http://dx.doi.org/10.1088/1126-6708/2005/07/076}{{\em JHEP} {\bfseries 07} (2005) 076}, \href{http://arxiv.org/abs/hep-th/0506093}{{\ttfamily arXiv:hep-th/0506093}}.

\bibitem{Metsaev:2005ar}
R.~R. Metsaev, ``{Cubic interaction vertices of massive and massless higher spin fields},'' \href{http://dx.doi.org/10.1016/j.nuclphysb.2006.10.002}{{\em Nucl. Phys. B} {\bfseries 759} (2006) 147--201}, \href{http://arxiv.org/abs/hep-th/0512342}{{\ttfamily arXiv:hep-th/0512342}}.

\bibitem{Metsaev:2007rn}
R.~R. Metsaev, ``{Cubic interaction vertices for fermionic and bosonic arbitrary spin fields},'' \href{http://dx.doi.org/10.1016/j.nuclphysb.2012.01.022}{{\em Nucl. Phys. B} {\bfseries 859} (2012) 13--69}, \href{http://arxiv.org/abs/0712.3526}{{\ttfamily arXiv:0712.3526 [hep-th]}}.

\bibitem{Metsaev:2022yvb}
R.~R. Metsaev, ``{Interacting massive and massless arbitrary spin fields in 4d flat space},'' \href{http://arxiv.org/abs/2206.13268}{{\ttfamily arXiv:2206.13268 [hep-th]}}.

\bibitem{Rarita:1941mf}
W.~Rarita and J.~Schwinger, ``{On a theory of particles with half integral spin},''
{\em Phys. Rev.} {\bfseries 60} (1941) 61.

\bibitem{Johnson:1960vt}
K.~Johnson and E.~C.~G. Sudarshan, ``{Inconsistency of the local field theory of charged spin 3/2 particles},''
\href{http://dx.doi.org/10.1016/0003-4916(61)90030-6}{{\em Annals Phys.} {\bfseries 13} (1961) 126--145}.

\bibitem{Deser:2000dz}
S.~Deser, V.~Pascalutsa, and A.~Waldron, ``{Massive spin 3/2 electrodynamics},'' \href{http://dx.doi.org/10.1103/PhysRevD.62.105031}{{\em Phys. Rev. D} {\bfseries 62} (2000) 105031}, \href{http://arxiv.org/abs/hep-th/0003011}{{\ttfamily arXiv:hep-th/0003011}}.

\bibitem{Deser:2001dt}
S.~Deser and A.~Waldron, ``{Inconsistencies of massive charged gravitating higher spins},'' \href{http://dx.doi.org/10.1016/S0550-3213(02)00199-2}{{\em Nucl. Phys. B} {\bfseries 631} (2002) 369--387}, \href{http://arxiv.org/abs/hep-th/0112182}{{\ttfamily arXiv:hep-th/0112182}}.

\bibitem{Buchbinder:2014hfa}
I.~L. Buchbinder, T.~V. Snegirev, and Y.~M. Zinoviev, ``{Formalism of gauge-invariant curvatures and constructing the cubic vertices for massive spin-$\frac{3}{2}$ field in AdS$_4$ space},'' \href{http://dx.doi.org/10.1140/epjc/s10052-014-3153-3}{{\em Eur. Phys. J. C} {\bfseries 74} no.~11, (2014) 3153}, \href{http://arxiv.org/abs/1405.7781}{{\ttfamily arXiv:1405.7781 [hep-th]}}.

\bibitem{Khabarov:2021djm}
M.~V. Khabarov and Y.~M. Zinoviev, ``{On massive spin-3/2 in the Fradkin\textendash{}Vasiliev formalism},'' \href{http://dx.doi.org/10.1088/1361-6382/ac1c1e}{{\em Class. Quant. Grav.} {\bfseries 38} no.~19, (2021) 195012}, \href{http://arxiv.org/abs/2105.01325}{{\ttfamily arXiv:2105.01325 [hep-th]}}.

\bibitem{Benakli:2023aes}
K.~Benakli, C.~A. Daniel, and W.~Ke, ``{Spin-3/2 and spin-2 charged massive states in a constant electromagnetic background},'' \href{http://dx.doi.org/10.1007/JHEP03(2023)212}{{\em JHEP} {\bfseries 03} (2023) 212}, \href{http://arxiv.org/abs/2302.06630}{{\ttfamily arXiv:2302.06630 [hep-th]}}.

\bibitem{Benakli:2022edf}
K.~Benakli, C.~A. Daniel, and W.~Ke, ``{Open superstring first mass level effective Lagrangian: Massive spin-3/2 fields in an electromagnetic background},'' \href{http://dx.doi.org/10.1016/j.physletb.2023.137788}{{\em Phys. Lett. B} {\bfseries 839} (2023) 137788}, \href{http://arxiv.org/abs/2211.13691}{{\ttfamily arXiv:2211.13691 [hep-th]}}.

\bibitem{Benakli:2021jxs}
K.~Benakli, N.~Berkovits, C.~A. Daniel, and M.~Lize, ``{Higher-spin states of the superstring in an electromagnetic background},'' \href{http://dx.doi.org/10.1007/JHEP12(2021)112}{{\em JHEP} {\bfseries 12} (2021) 112}, \href{http://arxiv.org/abs/2110.07623}{{\ttfamily arXiv:2110.07623 [hep-th]}}.

\bibitem{Klishevich:1998sr}
S.~M. Klishevich, ``{Electromagnetic interaction of massive spin 3 state from string theory},'' \href{http://dx.doi.org/10.1142/S0217751X00000185}{{\em Int. J. Mod. Phys. A} {\bfseries 15} (2000) 395--411}, \href{http://arxiv.org/abs/hep-th/9805174}{{\ttfamily arXiv:hep-th/9805174}}.

\bibitem{Klishevich:1998ng}
S.~M. Klishevich, ``{Massive fields of arbitrary integer spin in homogeneous electromagnetic field},'' \href{http://dx.doi.org/10.1142/S0217751X00000264}{{\em Int. J. Mod. Phys. A} {\bfseries 15} (2000) 535}, \href{http://arxiv.org/abs/hep-th/9810228}{{\ttfamily arXiv:hep-th/9810228}}.

\bibitem{Klishevich:1998yt}
S.~M. Klishevich, ``{Massive fields of arbitrary half integer spin in constant electromagnetic field},'' \href{http://dx.doi.org/10.1142/S0217751X00000306}{{\em Int. J. Mod. Phys. A} {\bfseries 15} (2000) 609--624}, \href{http://arxiv.org/abs/hep-th/9811030}{{\ttfamily arXiv:hep-th/9811030}}.

\bibitem{Porrati:2010hm}
M.~Porrati, R.~Rahman, and A.~Sagnotti, ``{String Theory and The Velo-Zwanziger Problem},'' \href{http://dx.doi.org/10.1016/j.nuclphysb.2011.01.007}{{\em Nucl. Phys.} {\bfseries B846} (2011) 250--282},
\href{http://arxiv.org/abs/1011.6411}{{\ttfamily arXiv:1011.6411 [hep-th]}}.

\bibitem{Buchbinder:2015uea}
I.~L. Buchbinder, V.~A. Krykhtin, and M.~Tsulaia, ``{Lagrangian formulation of massive fermionic higher spin fields on a constant electromagnetic background},'' \href{http://dx.doi.org/10.1016/j.nuclphysb.2015.04.008}{{\em Nucl. Phys. B} {\bfseries 896} (2015) 1--18}, \href{http://arxiv.org/abs/1501.03278}{{\ttfamily arXiv:1501.03278 [hep-th]}}.

\bibitem{Argyres:1989qr}
P.~C. Argyres and C.~R. Nappi, ``{Spin 1 Effective Actions From Open Strings},'' \href{http://dx.doi.org/10.1016/0550-3213(90)90305-W}{{\em Nucl. Phys. B} {\bfseries 330} (1990) 151--173}.

\bibitem{Argyres:1989cu}
P.~C. Argyres and C.~R. Nappi, ``{Massive Spin-2 Bosonic String States in an Electromagnetic Background},'' \href{http://dx.doi.org/10.1016/0370-2693(89)91055-1}{{\em Phys. Lett. B} {\bfseries 224} (1989) 89--96}.

\bibitem{Vines:2017hyw}
J.~Vines, ``{Scattering of two spinning black holes in post-Minkowskian gravity, to all orders in spin, and effective-one-body mappings},'' \href{http://dx.doi.org/10.1088/1361-6382/aaa3a8}{{\em Class. Quant. Grav.} {\bfseries 35} no.~8, (2018) 084002},
\href{http://arxiv.org/abs/1709.06016}{{\ttfamily arXiv:1709.06016 [gr-qc]}}.

\bibitem{Arkani-Hamed:2017jhn}
N.~Arkani-Hamed, T.-C. Huang, and Y.-t. Huang, ``{Scattering amplitudes for all masses and spins},'' \href{http://dx.doi.org/10.1007/JHEP11(2021)070}{{\em JHEP} {\bfseries 11} (2021) 070}, \href{http://arxiv.org/abs/1709.04891}{{\ttfamily arXiv:1709.04891 [hep-th]}}.

\bibitem{Guevara:2018wpp}
A.~Guevara, A.~Ochirov, and J.~Vines, ``{Scattering of Spinning Black Holes from Exponentiated Soft Factors},'' \href{http://dx.doi.org/10.1007/JHEP09(2019)056}{{\em JHEP} {\bfseries 09} (2019) 056},
\href{http://arxiv.org/abs/1812.06895}{{\ttfamily arXiv:1812.06895 [hep-th]}}.

\bibitem{Chung:2018kqs}
M.-Z. Chung, Y.-T. Huang, J.-W. Kim, and S.~Lee, ``{The simplest massive S-matrix: from minimal coupling to Black Holes},'' \href{http://dx.doi.org/10.1007/JHEP04(2019)156}{{\em JHEP} {\bfseries 04} (2019) 156},
\href{http://arxiv.org/abs/1812.08752}{{\ttfamily arXiv:1812.08752 [hep-th]}}.

\bibitem{Skvortsov:2023jbn}
E.~Skvortsov and M.~Tsulaia, ``{Cubic action for Spinning Black Holes from massive higher-spin gauge symmetry},'' \href{http://arxiv.org/abs/2312.08184}{{\ttfamily arXiv:2312.08184 [hep-th]}}.

\bibitem{Cangemi:2023bpe}
L.~Cangemi, M.~Chiodaroli, H.~Johansson, A.~Ochirov, P.~Pichini, and E.~Skvortsov, ``{Compton Amplitude for Rotating Black Hole from QFT},'' \href{http://arxiv.org/abs/2312.14913}{{\ttfamily arXiv:2312.14913 [hep-th]}}.

\bibitem{Monteiro:2014cda}
R.~Monteiro, D.~O'Connell, and C.~D. White, ``{Black holes and the double copy},'' \href{http://dx.doi.org/10.1007/JHEP12(2014)056}{{\em JHEP} {\bfseries 1412} (2014) 056},
\href{http://arxiv.org/abs/1410.0239}{{\ttfamily arXiv:1410.0239 [hep-th]}}.

\bibitem{Arkani-Hamed:2019ymq}
N.~Arkani-Hamed, Y.-t. Huang, and D.~O'Connell, ``{Kerr black holes as elementary particles},'' \href{http://dx.doi.org/10.1007/JHEP01(2020)046}{{\em JHEP} {\bfseries 01} (2020) 046},
\href{http://arxiv.org/abs/1906.10100}{{\ttfamily arXiv:1906.10100 [hep-th]}}.

\bibitem{Guevara:2020xjx}
A.~Guevara, B.~Maybee, A.~Ochirov, D.~O'Connell, and J.~Vines, ``{A worldsheet for Kerr},'' \href{http://dx.doi.org/10.1007/JHEP03(2021)201}{{\em JHEP} {\bfseries 03} (2021) 201}, \href{http://arxiv.org/abs/2012.11570}{{\ttfamily arXiv:2012.11570 [hep-th]}}.

\bibitem{Bern:2010ue}
Z.~Bern, J.~J.~M. Carrasco, and H.~Johansson, ``{Perturbative Quantum Gravity as a Double Copy of Gauge Theory},'' \href{http://dx.doi.org/10.1103/PhysRevLett.105.061602}{{\em Phys.Rev.Lett.} {\bfseries 105} (2010) 061602},
\href{http://arxiv.org/abs/1004.0476}{{\ttfamily arXiv:1004.0476 [hep-th]}}.

\bibitem{Porrati:2009bs}
M.~Porrati and R.~Rahman, ``{Causal Propagation of a Charged Spin 3/2 Field in an External Electromagnetic Background},'' \href{http://dx.doi.org/10.1103/PhysRevD.80.025009}{{\em Phys. Rev.} {\bfseries D80} (2009) 025009},
\href{http://arxiv.org/abs/0906.1432}{{\ttfamily arXiv:0906.1432 [hep-th]}}.

\bibitem{Cangemi:2023ysz}
L.~Cangemi, M.~Chiodaroli, H.~Johansson, A.~Ochirov, P.~Pichini, and E.~Skvortsov, ``{From higher-spin gauge interactions to Compton amplitudes for root-Kerr},'' \href{http://arxiv.org/abs/2311.14668}{{\ttfamily arXiv:2311.14668 [hep-th]}}.

\bibitem{Chiodaroli:2021eug}
M.~Chiodaroli, H.~Johansson, and P.~Pichini, ``{Compton black-hole scattering for s \ensuremath{\leq} 5/2},'' \href{http://dx.doi.org/10.1007/JHEP02(2022)156}{{\em JHEP} {\bfseries 02} (2022) 156}, \href{http://arxiv.org/abs/2107.14779}{{\ttfamily arXiv:2107.14779 [hep-th]}}.

\bibitem{Cangemi:2022bew}
L.~Cangemi, M.~Chiodaroli, H.~Johansson, A.~Ochirov, P.~Pichini, and E.~Skvortsov, ``{Kerr Black Holes From Massive Higher-Spin Gauge Symmetry},'' \href{http://dx.doi.org/10.1103/PhysRevLett.131.221401}{{\em Phys. Rev. Lett.} {\bfseries 131} no.~22, (2023) 221401}, \href{http://arxiv.org/abs/2212.06120}{{\ttfamily arXiv:2212.06120 [hep-th]}}.

\bibitem{Skvortsov:2022unu}
E.~Skvortsov and R.~Van~Dongen, ``{Minimal models of field theories: SDYM and SDGR},'' \href{http://dx.doi.org/10.1007/JHEP08(2022)083}{{\em JHEP} {\bfseries 08} (2022) 083}, \href{http://arxiv.org/abs/2204.09313}{{\ttfamily arXiv:2204.09313 [hep-th]}}.

\end{thebibliography}
\end{document}